\documentclass[reprint,aps,twocolumn,groupedaddress]{revtex4-1}
\usepackage{color}
\usepackage{graphicx}
\usepackage{threeparttable}
\usepackage{longtable}
\usepackage{enumerate}
\usepackage{amsfonts,amssymb}
\usepackage{amsmath}
\usepackage{stmaryrd}
\usepackage{dcolumn}
\usepackage{bm}
\usepackage{bbm}
\usepackage{algpseudocode}
\usepackage{algorithm}
\usepackage[all,cmtip]{xy}
\usepackage{url}

\bibpunct{}{}{,}{s}{}{\textsuperscript{,}}  

\newcommand{\tr}{\mathrm{Tr}}

\newcommand{\csch}{\mathrm{csch}}
\newcommand{\rint}{\mathrm{int}}
\newcommand{\bK}{\mathbf{K}}
\newcommand{\bI}{\mathbf{I}}
\newcommand{\bV}{\mathbf{V}}
\newcommand{\drm}{\mathrm{d}}
\newcommand{\Drm}{\mathrm{D}}

\begin{document}
\title{On the generalization of the exponential basis
for tensor network representations of long-range interactions in two and three dimensions}
\author{Zhendong Li}\email{zhendongli2008@gmail.com}
\affiliation{\footnotesize{Division of Chemistry and Chemical Engineering,
California Institute of Technology, Pasadena, CA 91125, USA}}
\affiliation{\footnotesize{Key Laboratory of Theoretical and Computational Photochemistry, Ministry of Education, College of Chemistry, Beijing Normal University, Beijing 100875, China}}
\author{Matthew J. O'Rourke}
\affiliation{\footnotesize{Division of Chemistry and Chemical Engineering,
California Institute of Technology, Pasadena, CA 91125, USA}}
\author{Garnet Kin-Lic Chan}\email{gkc1000@gmail.com}
\bigskip
\affiliation{\footnotesize{Division of Chemistry and Chemical Engineering,
    California Institute of Technology, Pasadena, CA 91125, USA}}
\bigskip
\date{\today}

\begin{abstract}
In one dimension (1D), a general decaying long-range interaction can be fit to a sum of exponential interactions $e^{-\lambda r_{ij}}$ with varying exponents $\lambda$, each of which can be represented by a simple matrix product operator (MPO) with bond dimension $D=3$. Using this technique,
efficient and accurate simulations of 1D quantum systems with long-range interactions can be performed
using matrix product states (MPS). However, the extension of this construction to higher dimensions is not obvious.
We report how to generalize the exponential basis to 2D and 3D by defining the basis functions as the Green's functions of the discretized Helmholtz equation for different Helmholtz parameters $\lambda$, a construction which is valid for lattices of any spatial dimension. Compact tensor network representations can then be found
for the discretized Green's functions, by expressing them as correlation functions of auxiliary fermionic fields with nearest neighbor interactions via Grassmann Gaussian integration. Interestingly, this analytic construction in 3D yields a $D=4$ tensor network representation of correlation functions which (asymptotically) decay as the inverse distance ($r^{-1}_{ij}$), thus generating the (screened) Coulomb potential on a cubic lattice. These
techniques will be useful in tensor network simulations of realistic materials.
\end{abstract}
\maketitle

\section{Introduction}
To understand the electronic properties of realistic materials, it is important to account for the effects of the long-range Coulomb interaction between electrons. However,
solving the many-electron Schr{\"o}dinger equation (SE) including
the Coulomb potential $V_{\rint}=\sum_{i<j}1/r_{ij}$
is challenging and typically involves uncontrolled approximations.
A promising approach 
where there is a systematic control of accuracy is provided by the density matrix renormalization group (DMRG) algorithm\cite{white1992dmrg,white1993dmrg}
and its higher-dimensional extensions such as projected entangled-pair states (PEPS)
\cite{nishino1996corner,verstraete2004renormalization,
verstraete2006criticality}, which reduce the effective dimensionality of the SE by exploiting the locality typically found in physical systems.
In fact, the DMRG has already been widely applied to solve the SE for
molecules\cite{white1999ab,chan2002highly,chan2016matrix} within a finite basis expansion.
However, to exploit the power of higher dimensional tensor network states (TNS)\cite{orus2014practical}, the standard orbital (or spectral) basis\cite{boyd2001chebyshev} expansion for the SE is not ideal. This is because whereas the variational freedom in the TNS parametrization scales only
linearly with system size $A$, the representation of the Coulomb operator has a large number
of terms that scales like $O(A^4)$.

In our earlier work\cite{o2018efficient}, we proposed to combine higher-dimensional TNS with
a real-space lattice discretization of the SE, which reformulates the SE as an extended Hubbard model
with density-density type long-range interactions $\hat{V}_{\rint}=\sum_{i<j}v_{ij}^{ee}n_{i}n_{j}$,
where $i,j$ label lattice sites, $v_{ij}^{ee}=1/r_{ij}$, $r_{ij}=|\mathbf{r}_i-\mathbf{r}_j|=|\vec{i}-\vec{j}|l$,
$l$ is the lattice spacing, and $n_i$ is the number operator.
In this form, the discretization error can be systematically controlled by reducing the lattice spacing $l$. The representation of the Hamiltonian
is also improved for TNS simulations, as there are now $O(A^2)$  interaction terms between
the electron sites.
Nonetheless, even with this reduction, a term-by-term evaluation of the Coulomb interaction energy $\langle\Psi|\hat{V}_{\mathrm{int}}|\Psi\rangle$ (in which the values of the potential are explicitly computed for each term) still leads to an undesirable
quadratic computational scaling with  system size.

In the one dimensional (1D) case, the above problem can be overcome by fitting the
Coulomb interaction $1/r_{ij}$ to a sum of exponentials
$\sum_{t=1}^{N_t}c_t e^{-\lambda_t r_{ij}}$\cite{crosswhite2008applying,pirvu2010matrix},
where the number of terms $N_t$ depends only on the target fitting
accuracy rather than the system size $A$. Each exponential interaction $\hat{V}=\sum_{i<j}e^{-\lambda r_{ij}}n_in_j$ can then be
represented by a matrix product operator (MPO)\cite{crosswhite2008applying,
crosswhite2008finite,pirvu2010matrix,verstraete2004matrix,
mcculloch2007density,schollwock2011density} with bond dimension $D=3$,
\begin{gather}
\hat{V} =(\hat{W}[1]\hat{W}[2]\cdots \hat{W}[N])_{11},\nonumber\\
\hat{W}[i] = \left[
\begin{array}{ccc}
I & e^{-\lambda l/2} n_{i} & 0 \\
0 & e^{-\lambda l} I & e^{-\lambda l/2} n_{i} \\
0 & 0 & I \\
\end{array}\right].\label{expMPO}
\end{gather}
In this way, computing $\langle\Psi|\hat{V}_{\rint}|\Psi\rangle$ with
$|\Psi\rangle$ represented by a matrix product state (MPS)\cite{schollwock2011density}
with bond dimension $D$ scales as $O(N_tAD^3)$, that is, linearly with the system size.
In combination with DMRG, this representation has been used to simulate interacting 1D
models in the continuum limit\cite{stoudenmire2012one,wagner2012reference,
slicedbasis,dolfi2012multigrid} with controllable accuracy by systematically reducing
the spacing $l$ and increasing the bond dimension $D$.

Generalizing such a construction for long-range interactions to higher dimensions is, however, nontrivial.
In 2D or 3D, the exponential $e^{-\lambda r_{ij}}$
cannot be formed as a product of weight factors $e^{-\lambda l}$ along the path from $i$ to $j$, as it is done in 1D.
In our previous work\cite{o2018efficient} on the 2D case,
we used the spin-spin correlation functions of
the 2D classical Ising model $\langle \sigma_i\sigma_j\rangle_{\beta_t}$ (where $\sigma_i,\sigma_j\in\{+1,-1\}$ and $\beta_t$ is the inverse temperature)
as a basis to numerically fit the Coulomb interaction on a square lattice to
 a sum of the correlation functions $\langle \sigma_i\sigma_j\rangle_{\beta_t}$ at different temperatures,
viz., $1/r_{ij}\approx \sum_{t=1}^{N_t}c_t\langle \sigma_i\sigma_j\rangle_{\beta_t}$.
However, an important difference between the 1D and 2D cases is that
$\langle \sigma_i\sigma_j\rangle_{\beta_t}$ at short lattice distances is not a smooth and radially isotropic function of $r_{ij}$.
To control these  errors in 2D, we embedded the physical lattice into a larger underlying Ising lattice, for details, see Ref. \cite{o2018efficient}. [NB: In the experimental setting, a related recent proposal uses auxiliary particles
  on a larger underlying lattice to mediate the Coulomb interaction between physical particles for
  the purposes of building an analog quantum simulation of Hamiltonians with long-range interactions\cite{arguello2018analog}.]
Then, since the Ising correlation function $\langle \sigma_i\sigma_j\rangle_{\beta_t}$ can be represented by a classical PEPS with $D=2$\cite{verstraete2006criticality,zhao2010renormalization},
by using the finite automata construction\cite{crosswhite2008finite,pirvu2010matrix,frowis2010tensor} to couple operators $n_in_j$ with weights $\langle\sigma_i\sigma_j\rangle_{\beta_t}$,
the long-range interaction $\sum_{i<j}\langle \sigma_i\sigma_j\rangle_{\beta_t} n_in_j$ can be represented by a projected entangled-pair operator (PEPO) with maximal bond dimension $D=2\times 3=6$ or $D=2\times 4=8$, depending on the choice of finite automata rules\cite{o2018efficient}
to represent the operator sum $\sum_{i<j}n_in_j$.
Consequently, the long-range interaction $\hat{V}_{\rint}$ can be approximated as a sum of PEPOs with constant bond dimension, which is
analogous to the 1D case.

Nonetheless, using a numerically defined basis $\langle \sigma_i\sigma_j\rangle_{\beta_t}$ to expand the interaction in 2D complicates matters
significantly as compared to using the analytic basis $e^{-\lambda_t r_{ij}}$ in 1D.
For example, in Ref.~\cite{o2018efficient} the performance of the fit in various limits could only be assessed numerically, and the
analysis was restricted to the 2D square lattice.
In this work, we define an analytic framework which contains
the exponential basis construction in 1D and provides
a natural generalization to lattices in any spatial dimension, although
we will focus explicitly only on 2D and 3D. Importantly, this formulation produces a set of
long-range basis functions with
explicit tensor network (TN) representations with small, constant bond dimensions
such that decaying long-range interactions can be
approximated by a sum of tensor network operators (TNO) efficiently.

Specifically, in Sec. \ref{sec:Basis}, we introduce the framework by
defining appropriate basis functions (in any dimension) as the Green's function of the discretized Helmholtz equation.
Then, using Grassmann Gaussian integration, the discretized Green's functions can be expressed as correlation functions of auxiliary fermionic fields $\langle c_i\bar{c}_j\rangle$. This allows us to show in what sense the 1D geometry is special: there are fundamental differences between exponentials in 1D and their higher dimensional extensions. In 1D, both the discretization and finite size errors can be removed analytically,
resulting in a TN representation of the continuum exponential function,
while in 2D and 3D they cannot be removed in a way compatible with
a simple TN representation of low bond dimension. In Sec. \ref{sec:1D}, the MPO representation in Eq. \eqref{expMPO}
is re-derived within the proposed framework. In Sec. \ref{sec:2D} and \ref{sec:3D}, we show that the discretized Green's functions in 2D and 3D, respectively, also have very compact TN representations. Most interestingly, the analytic construction in 3D yields a TN representation of correlation functions decaying as $r^{-1}_{ij}$ asymptotically, which generates the (screened) Coulomb potential on the cubic lattice.
The subroutines for constructing TN representations and examples for numerical contractions of the resulting TN are available online\cite{linkToLSnD}.
Finally, conclusions are drawn in Sec. \ref{sec:Conclusion}.

\section{Fitting basis in any dimension}\label{sec:Basis}

One way to view the exponential $e^{-\lambda|\mathbf{r}-\mathbf{r}'|}$ in 1D is that it is the Green's function of the Helmholtz equation in free space (up to a constant $\frac{1}{2\lambda}$),
\begin{eqnarray}
(-\nabla ^{2}+\lambda^{2})G(\mathbf{r},\mathbf{r}')=\delta(\mathbf{r}-\mathbf{r}'),\label{Helmholtz}
\end{eqnarray}
or in other words, it is the Fourier transform of the momentum-space kernel $(|\mathbf{k}|^2+\lambda^{2})^{-1}$.
In 2D and 3D, the solution to Eq. \eqref{Helmholtz} is given by the modified Bessel function $\frac{1}{2\pi}K_{0}(\lambda |\mathbf{r}-\mathbf{r}'|)$ and the screened Coulomb potential
$\frac{e^{-\lambda|\mathbf{r}-\mathbf{r}'|}}{4\pi |\mathbf{r}-\mathbf{r}'|}$, respectively, both of
which decay exponentially at large distance for $\lambda>0$.
Therefore, similarly to in the 1D case, if we are able to find the corresponding TN representation of these interactions, we can use them as
a basis of functions with which to represent decaying long-range interactions.

\begin{figure}[ht]
\centering
\begin{tabular}{cc}
\multicolumn{2}{c}{\includegraphics[width=0.3\textwidth]{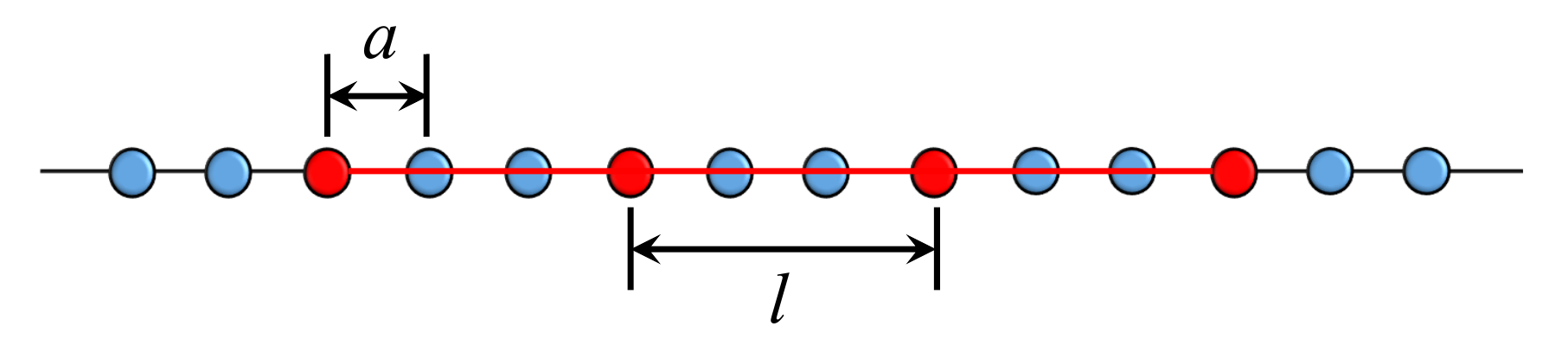}} \\
\multicolumn{2}{c}{(a) 1D infinite lattice} \\
\includegraphics[width=0.2\textwidth]{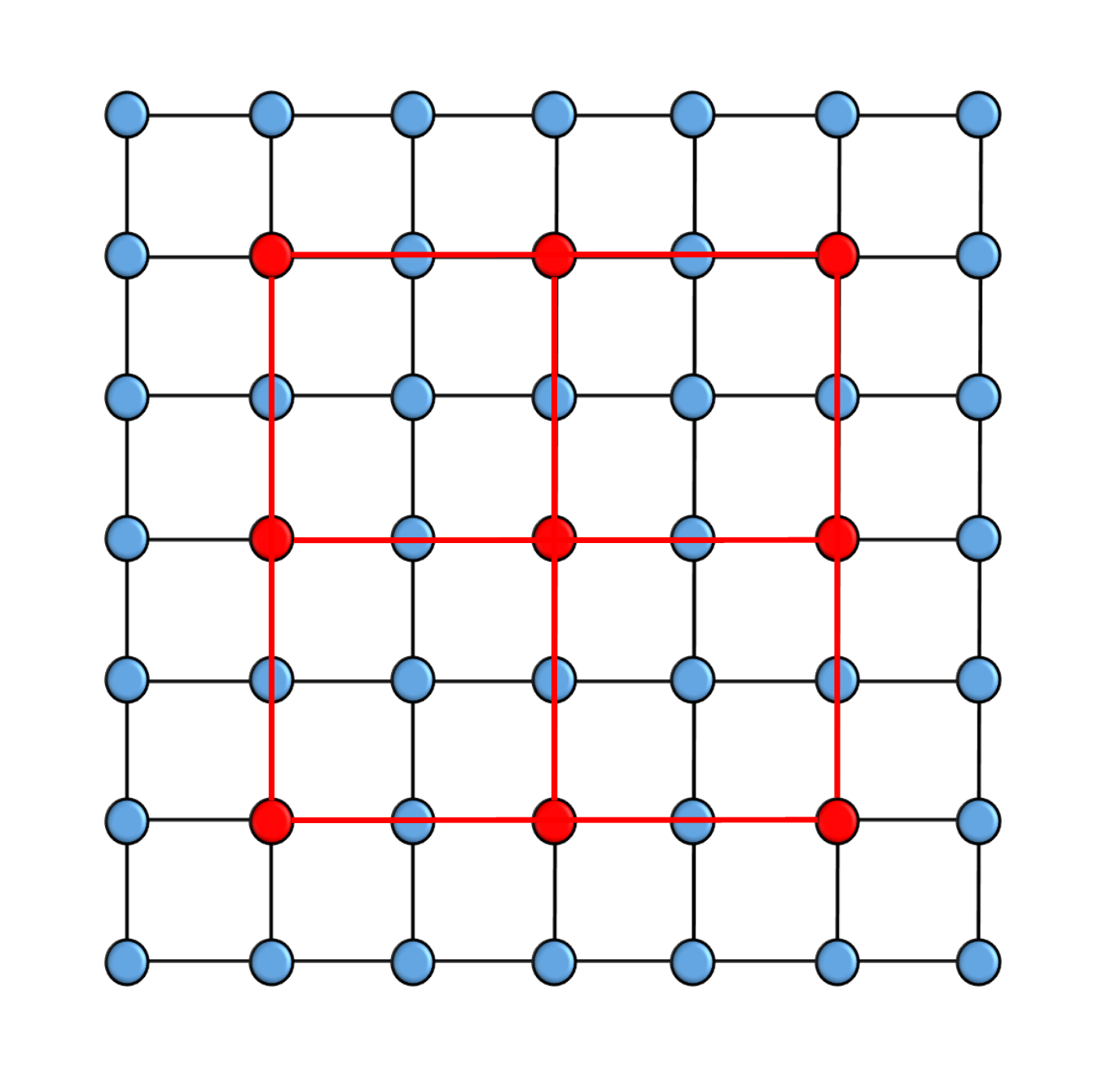} &
\includegraphics[width=0.24\textwidth]{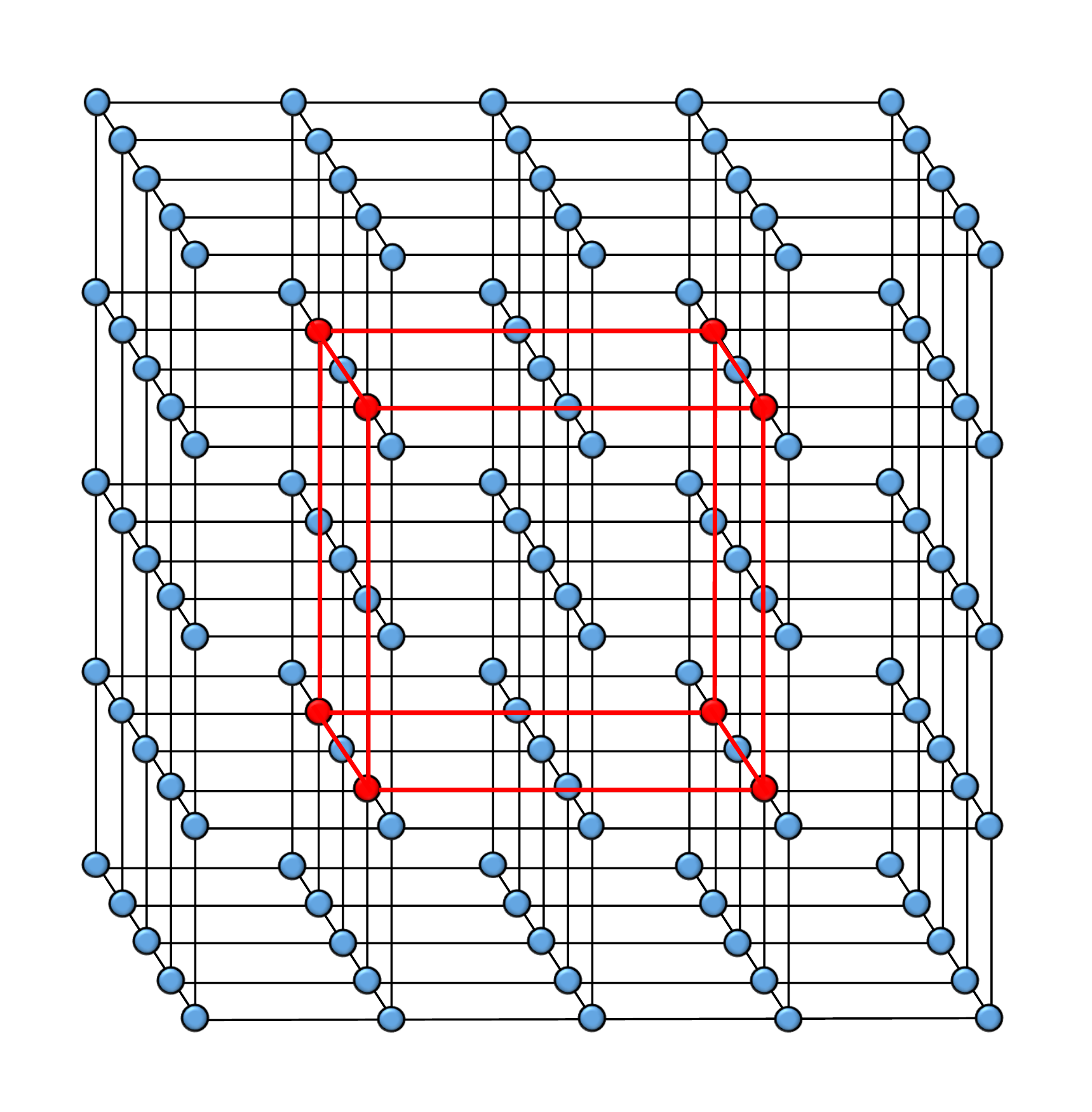} \\
\multicolumn{2}{c}{(b) 2D and 3D lattices} \\
\multicolumn{2}{c}{\includegraphics[width=0.4\textwidth]{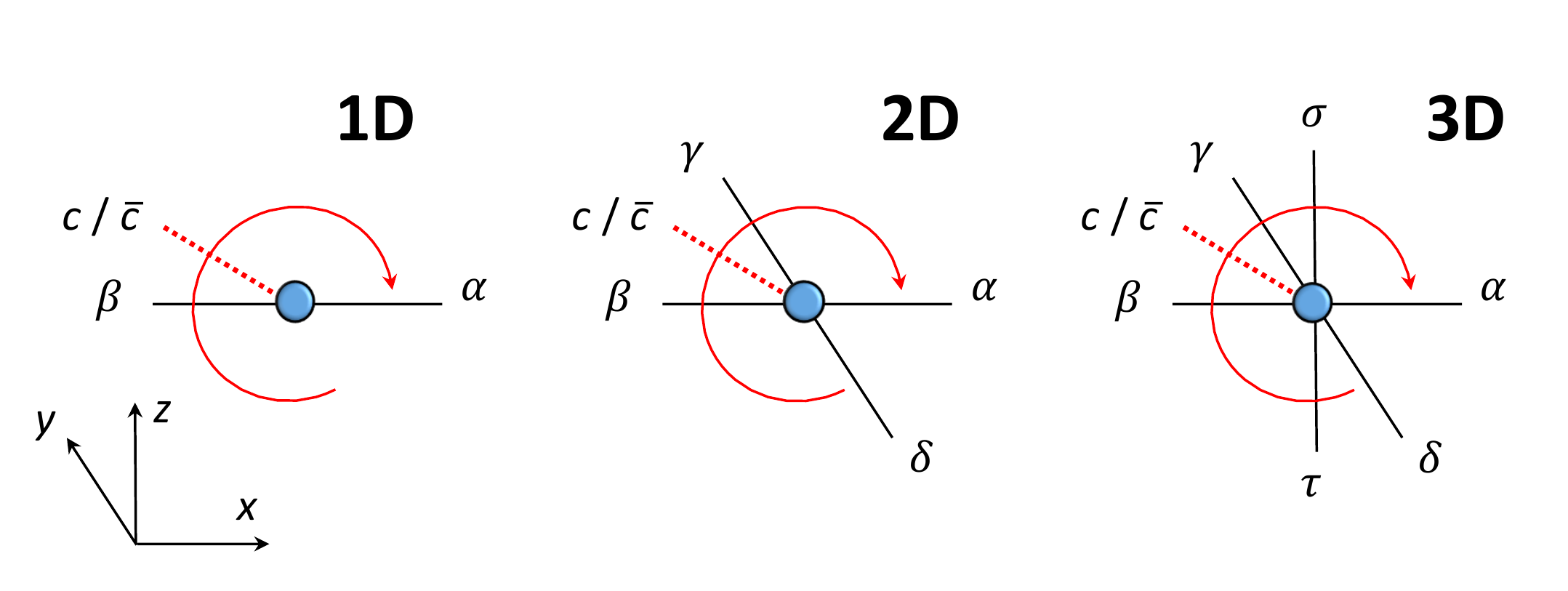}} \\
\multicolumn{2}{c}{(c) Graphical representation of local tensors} \\
\end{tabular}
\caption{Lattice discretization in $d$-dimensional case (a,b)
and rules for writing down local tensors (c)
for representing $V_{ij}$ in Eq. \eqref{GrassmannIntegration}.
The blue dots refer to the underlying discretized lattice with spacing $a$ for mediating
long-range interactions, and the red dots refer to the physical lattice with spacing $l$.}\label{fig:1d}
\end{figure}

To begin, we consider the discretized version of Eq. \eqref{Helmholtz} on a $d$-dimensional cubic lattice with spacing $a$ (see
Figures \ref{fig:1d}(a) and \ref{fig:1d}(b)),
\begin{eqnarray}
(\bK_d+\lambda^{2}a^2 \bI)\bV=\bI,\label{dGF}
\end{eqnarray}
where $\bK_d/a^2$ is the discretized version of $(-\nabla^2)$, which
in the simplest case can be represented by the central difference scheme with open-boundary conditions (OBC),
\begin{eqnarray}
\bK_1 = \left[\begin{array}{cccccc}
2  & -1 &  0 & \cdots & 0 & 0\\
-1 & 2  & -1 & \cdots & 0 & 0\\
0  & -1 &  2 & \cdots & 0 & 0\\
\vdots & \vdots & \vdots & \ddots & \vdots & \vdots \\
0  & 0  &  0 & \cdots & 2 & -1\\
0  & 0  &  0 & \cdots & -1 & 2\\
\end{array}\right]_{N\times N}.\label{Kbare}
\end{eqnarray}
Here the length of the lattice is $L=(N+1)a$, $\bK_2=\bK_1\otimes\bI+\bI\otimes\bK_1$, and
$\bK_3=\bK_1\otimes\bI\otimes\bI+\bI\otimes\bK_1\otimes\bI+
\bI\otimes\bI\otimes\bK_1$. The matrix $V_{ij}$ is related to the
continuum Green's function by
\begin{eqnarray}
G(\mathbf{r}_i,\mathbf{r}_j)&=&\lim_{a\rightarrow 0}a^{-d}[(\bK_d/a^2+\lambda^{2}\bI)^{-1}]_{
i=\frac{\mathbf{r}_i}{a},j=\frac{\mathbf{r}_j}{a}}\nonumber\\
&=&\lim_{a\rightarrow 0}a^{-d+2}V_{i=\frac{\mathbf{r}_i}{a},j=\frac{\mathbf{r}_j}{a}},\label{GxyVij}
\end{eqnarray}
where the factor $a^{-d}$ in the first identity comes from scaling
\footnote{The factor $a^{-d}$ can be understood by considering the discretization
of the Gaussian functional integral for
$G(\mathbf{r}_i,\mathbf{r}_j)=\frac{\int \Drm[\phi]e^{-S}\phi(\mathbf{r}_i)\phi(\mathbf{r}_j)}
{\int \Drm[\phi]e^{-S}}$ with $S=\frac{1}{2}\int \drm^d\mathbf{r}
\phi(\mathbf{r})(-\nabla^2+\lambda^2)\phi(\mathbf{r})$.
On a $d$-dimensional lattice with spacing $a$, $S$ becomes
$S=\frac{1}{2}\phi_i(K_{ij}/a^2+\lambda^2\delta_{ij})\phi_j a^d$
such that by a change of variable $\tilde{\phi}_i=\phi_i a^{d/2}$,
$G(\mathbf{r}_i,\mathbf{r}_j)=
\frac{\int \prod_k\drm\phi_k e^{-S}\phi_i\phi_j}{\int \prod_k\drm\phi_k e^{-S}}=
\frac{\int \prod_k\drm\tilde{\phi}_k e^{-1/2 \tilde{\phi}_i(K_{ij}/a^2+\lambda^2\delta_{ij})\tilde{\phi}_j}
\tilde{\phi}_i\tilde{\phi}_ja^{-d}}
{\int \prod_k\drm\tilde{\phi}_k e^{-1/2\tilde{\phi}_i(K_{ij}/a^2+\lambda^2\delta_{ij})\tilde{\phi}_j}}
=a^{-d}[(\bK_d/a^2+\lambda^{2}\bI)^{-1}]_{ij}$.}.
In order to analytically relate $V_{ij}$ to a tensor network, we can use Gaussian integration to express $V_{ij}$ as the correlation function
of an auxiliary bosonic or fermionic system with nearest neighbor couplings.
We will use Grassmann variables such that $V_{ij}$ can be written as
\begin{eqnarray}
V_{ij}&=&\frac{1}{Z}\int \Drm[\bar{c},c] e^{-\bar{c}^T (\bK_d+\lambda^{2}a^2\bI) c}(c_i\bar{c}_j)\triangleq
\langle c_i\bar{c}_j\rangle,\nonumber\\
Z &=& \int \Drm[\bar{c},c] e^{-\bar{c}^T (\bK_d+\lambda^{2}a^2\bI) c},\label{GrassmannIntegration}
\end{eqnarray}
where a pair of Grassmann variables $\{\bar{c}_i,c_i\}$ is associated
with each lattice site. The advantage of using auxiliary fermions
instead of bosons is that the resulting tensor network representations for $Z$ and $V_{ij}$ will have finite
as opposed to infinite\cite{janik2019exact} bond dimension (vide post).

The special nature of the 1D geometry can now be examined:
As depicted in Figure \ref{fig:1d}(a), the
physical lattice (red) with spacing $l$ can be placed
on an infinite underlying lattice with spacing $a$ with associated Grassmann variables at each site. The infinite boundary sites can then be integrated out analytically ($N\rightarrow\infty$ while keeping $a$ fixed) to remove boundary effects, and the same can be done for the infinite
interior sites ($a\rightarrow0$ while keeping $l$ fixed) in order to remove
the discretization error. We then obtain an effective matrix $\bK_1'$ defined only on the physical lattice that replaces $(\bK_1+\lambda^{2}a^2\bI)$ in Eq. \eqref{GrassmannIntegration},
\begin{gather}
\bK_1' = \left[\begin{array}{cccccc}
k_b & k_c & 0 & \cdots & 0 & 0 \\
k_c & k_i & k_c & \cdots & 0 & 0 \\
0 & k_c & k_i & \cdots & 0 & 0 \\
\vdots & \vdots & \vdots & \ddots & \vdots & \vdots \\
0  & 0  &  0 & \cdots & k_i & k_c \\
0  & 0  &  0 & \cdots & k_c & k_b \\
\end{array}\right]_{n\times n},\nonumber\\
k_b=\frac{1}{2}(1+\coth\xi),\;
k_c=-\frac{1}{2}\csch\xi,\;
k_i=\coth\xi,\;
\xi=\lambda l.\label{Keff}
\end{gather}
It can be easily verified that the inverse of $\bK_1'$ indeed gives the exponential interaction
on the $n$-site physical lattice with spacing $l$, i.e., $(\bK_1')^{-1}_{ij}=e^{-\xi|i-j|}$.
However, in 2D and 3D (as shown in Figure \ref{fig:1d}(b)), integrating out the boundary and interior sites to obtain  continuum limit interactions between the physical sites
introduces couplings among all the physical sites. This results in a dense matrix $\bK_d'$,
which does not lead to a simple exact TN representation with constant bond dimension, because every tensor would require a bond to every other tensor.
Therefore, in the following discussion, we will focus on finding a TN representation
for the discrete analog $V_{ij}$ \eqref{GrassmannIntegration} of the
continuum Green's function in 2D and 3D.
Once this is done, the discretization error and finite size error in representing continuum interactions can be reduced by choosing a suitably large or infinite underlying Grassmann lattice and embedding the physical sites (red) in it (blue) as shown in Figure \ref{fig:1d}(b)
to effectively work with a smaller lattice spacing $a$ in $a^{-d+2}V_{i=\frac{\mathbf{r}_i}{a},j=\frac{\mathbf{r}_j}{a}}$ \eqref{GxyVij}, which is  similar in spirit to our previous work\cite{o2018efficient}
where we used an underlying larger Ising lattice to mediate interactions.
Interestingly, this construction in 3D  yields a TN representation of correlation functions that decays as $r^{-1}_{ij}$ asymptotically for $\lambda=0$.

In order to explicitly represent $V_{ij}$ as a tensor network, we note that the partition function $Z$ introduced in Eq. \eqref{GrassmannIntegration} is similar to that of the Ising model, which is easily written as a TN \cite{verstraete2006criticality,zhao2010renormalization}
in any dimension. Here, however, Grassmann variables are used rather than the spins $\sigma_i\in\{+1,-1\}$, and Grassmann integration
replaces the summation over spins.
Similarly to in the TN representation of the Ising $Z$, by factorizing $e^{\beta\sigma_{i}\sigma_j}$ into local quantities coupled by a virtual bond\cite{o2018efficient},
we can rewrite the nearest neighbor
coupling term in Eq. \eqref{GrassmannIntegration} as
\begin{eqnarray}
e^{\bar{c}_i c_j+\bar{c}_j c_i}
=1+\bar{c}_i c_j+\bar{c}_j c_i+
\bar{c}_i c_j\bar{c}_j c_i=\sum_{m=1}^{4}\alpha_{i,m}\beta_{j,m}.\label{expDecomposition}
\end{eqnarray}
The termination of the series for the exponential in the first equality is due to the nilpotency of
Grassmann variables, which is the advantage of using fermionic rather than bosonic
Gaussian integration in \eqref{GrassmannIntegration}.
The decomposition in the second equality can be performed in various ways. For simplicity,
we use the following form for the local factors $\alpha_{i}$ and $\beta_{j}$,
\begin{eqnarray}
\alpha_i &=& (1,\bar{c}_i,c_i,\bar{c}_i c_i),\nonumber\\
\beta_j  &=& (1,c_j,-\bar{c}_j,-\bar{c}_j c_j).\label{expVector}
\end{eqnarray}
Thus, instead of  $D=2$ for the bond dimension of the TN representation of $Z$ for the Ising model, we will have a $D=4$ construction here. Now the partition function $Z$ \eqref{GrassmannIntegration} can be expressed as a product of local ``projectors'' and terms
for each bond between two sites, e.g., in 3D it reads
\begin{eqnarray}
Z&=&\int \Drm[\bar{c},c]
\prod_k Q_k\prod_{<i,j>} B^x_{ij}\prod_{<m,n>}B^y_{mn} \prod_{<p,q>}B^z_{pq},\label{Z3d}
\end{eqnarray}
where $Q_k=e^{-(\bK_{d}+\lambda^2a^2\mathbf{I})_{kk}\bar{c}_k c_k}=
1-(2d+\lambda^2a^2)\bar{c}_k c_k$ with $d=3$ here,
and $B^x_{ij}$, $B^y_{ij}$, and $B^z_{ij}$ represent
the decomposed pairs in Eq. \eqref{expDecomposition} in
different directions. To distinguish the
pairs in different directions, in the following discussion we will use different
pairs of Greek letters for different directions even though
they denote the same vectors as in Eq. \eqref{expVector}: $\alpha,\beta$ for pairs in the $X$ direction;
$\gamma,\delta$ for pairs in the $Y$ direction; and
$\sigma,\tau$ for pairs in the $Z$ directions, see Figure \ref{fig:1d}(c).
That is, $B_{ij}^x=\alpha_i\beta_j$, $B_{ij}^y=\gamma_i\delta_j$,
and $B_{ij}^z=\sigma_i\tau_j$, where the summations over components have been
omitted for simplicity.

The partition function for the Ising model can be written in the same form as
Eq. \eqref{Z3d}, and by collecting the local factors belonging to
the same site together, $Z$ can be represented as a tensor network,
and the same strategy applies for the correlation function $\langle\sigma_{i}\sigma_{j}\rangle$.
However, in our case, due to the anti-commuting property of Grassmann
variables, additional sign factors will appear in moving variables in Eq. \eqref{Z3d}
to their respective local site. We note that Eq. \eqref{Z3d} for 2D is structurally similar to the fermionic PEPS (fPEPS)
\cite{kraus2010fermionic,pivzorn2010fermionic}, and it can be viewed as
a ``classical'' fPEPS without a physical index. In Sec. \ref{sec:2D} and Sec. \ref{sec:3D} we will show how to
express $Z$ and $V_{ij}$ as TN in 2D and 3D, and in particular,
how to deal with the sign factors that appear in different dimensions
by developing graphical rules similar to that for fPEPS\cite{corboz2010simulation}.
However, before this, we will first show how the present construction in 1D reproduces
the MPO in Eq. \eqref{expMPO}.

\section{Revisiting the 1D MPO representation}\label{sec:1D}
An important simplification in 1D is that the product
$\prod_{<i,j>} B^x_{ij}$ in  Eq. \eqref{Z3d}
is already in the desired form, viz.,
$\prod_{<i,j>} B^x_{ij}=\prod_{i=1}^{N-1}
(\alpha_{i}\beta_{i+1})=\alpha_{1}
\prod_{i=2}^{N-1}(\beta_{i}\alpha_{i})\beta_{N}$,
where the subscripts for components in $\alpha$ and $\beta$ have been omitted for simplicity.
Similarly, for correlation functions $\langle c_i\bar{c}_j\rangle$ ($i<j$),
the necessary product $(\prod_{<k,l>} B_{kl}^{x})(c_i\bar{c}_{j})$ can be arranged into
local products $\alpha_{1}(\beta_{2}\alpha_{2})\cdots
(\beta_{i} c_i\alpha_{i})\cdots
(\beta_{j} \bar{c}_{j}\alpha_{j})\cdots(\beta_{N-1}\alpha_{N-1})\beta_{N}$
without introducing any sign factors, because in moving
$c_i$ or $\bar{c}_j$, the terms $(\alpha_{i}\beta_{i+1})$ that must be passed over correspond
to a bond and are always of even parity, i.e., products of an even number of Grassmann variables.
Therefore, by defining the following local tensors for $\bK_1$ \eqref{Kbare},
\begin{eqnarray}
(A_{k})_{lr}&=& \int \drm \bar{c}_k\drm c_k\; Q_k\beta_{k,l}\alpha_{k,r}
=
\left[\begin{array}{cccc}
(2d+\lambda^2a^2) & 0 & 0 & -1 \\
0 & 1 & 0 & 0 \\
0 & 0 & 1 & 0 \\
1 & 0 & 0 & 0
\end{array}\right],\nonumber\\
(B_{k})_{lr} &=& \int \drm \bar{c}_k\drm c_k\; Q_k\beta_{k,l}(c_k)\alpha_{k,r}
=\left[\begin{array}{cccc}
0 & 1 & 0 & 0 \\
0 & 0 & 0 & 0 \\
1 & 0 & 0 & 0 \\
0 & 0 & 0 & 0 \\
\end{array}\right],\nonumber\\
(C_{k})_{lr} &=& \int \drm \bar{c}_k\drm c_k\; Q_k\beta_{k,l}(\bar{c}_k)\alpha_{k,r}
=\left[\begin{array}{cccc}
0 & 0 & -1 & 0 \\
1 & 0 & 0 & 0 \\
0 & 0 & 0 & 0 \\
0 & 0 & 0 & 0 \\
\end{array}\right],\label{localTensor1D}
\end{eqnarray}
the correlation function $V_{ij}$ ($i<j$) can be written as
\begin{eqnarray}
V_{ij}=\frac{1}{Z}(A_1\cdots B_i\cdots C_j \cdots A_{N})_{11},\;\; Z = (A_1\cdots A_{N})_{11}.\label{V1D}
\end{eqnarray}
When coupled with the operators $n_in_j$ via the finite
automata construction\cite{frowis2010tensor,pirvu2010matrix,crosswhite2008finite}, this form of $V_{ij}$ gives an
MPO representation for $\sum_{i<j}V_{ij}n_in_j$ with
bond dimension $D=3\times 4=12$,
that is, the factor $\hat{W}[i]$ appearing in the analog of Eq. \eqref{expMPO} reads
\begin{gather}
\hat{W}[i]
 = \left[
\begin{array}{ccc}
A_i\otimes I & B_{i}\otimes n_{i} & 0 \\
0 & A_i\otimes I & C_i\otimes n_{i} \\
0 & 0 & A_i\otimes I \\
\end{array}\right].\label{expMPO2}
\end{gather}
This MPO construction for 1D is not optimal in the sense that
it is very sparse and compressible in view of the sparse structure of $B_k$ and $C_k$.
We illustrate this compression for $\bK_1'$ \eqref{Keff} in order
to eventually recover the MPO in Eq. \eqref{expMPO}.

In this case, we can replace $\bar{c}$ in Eq. \eqref{expVector} by $-k_c\bar{c}$
to factorize $e^{-k_c(\bar{c}_ic_j+\bar{c}_jc_i)}$. Then,
the local tensors read
\begin{eqnarray}
(A_{k})_{lr}&=&
\left[\begin{array}{cccc}
k_{b,i} & 0 & 0 & k_c \\
0 & -k_c & 0 & 0 \\
0 & 0 & -k_c & 0 \\
-k_c & 0 & 0 & 0
\end{array}\right],\nonumber\\
(B_{k})_{lr} &=&
\left[\begin{array}{cccc}
0 & -k_c & 0 & 0 \\
0 & 0 & 0 & 0 \\
-k_c & 0 & 0 & 0 \\
0 & 0 & 0 & 0 \\
\end{array}\right],\nonumber\\
(C_{k})_{lr} &=&
\left[\begin{array}{cccc}
0 & 0 & -1 & 0 \\
1 & 0 & 0 & 0 \\
0 & 0 & 0 & 0 \\
0 & 0 & 0 & 0 \\
\end{array}\right],\label{localTensor1DKp}
\end{eqnarray}
where $(A_k)_{11}=k_i$ ($k_b$) for interior (boundary) sites on a lattice with $n$ sites.
It can be found that the products $(A_1\cdots A_{i-1})_{11}=(\frac{1}{2}\csch\xi e^\xi)^{i-1}$
($2\le i\le n$), $(A_{j+1}\cdots A_n)_{11}=(\frac{1}{2}\csch\xi e^\xi)^{n-j}$ ($1\le j\le n-2$), and $Z=(A_1\cdots A_{n})_{11}=(\frac{1}{2}\csch\xi e^\xi)^{n-1}$.
Due to the sparsity of the local tensors, we can observe that for $V_{ij}$ \eqref{V1D},
only the element $(B_{i})_{12}$ can contribute and $(B_i)_{31}$ cannot,
because at the boundary
$(A_1)_{13}=0$ such that $(A_1 \ldots A_{i-1})_{13}=0$.
Similar observations apply to $C_{j}$.
Thus, $V_{ij}$ in Eq. \eqref{V1D} can be rewritten as
\begin{eqnarray}
V_{ij}
&=&
\frac{1}{Z}(A_1\cdots A_{i-1})_{11}(B_i)_{12}(A_{i+1}\cdots A_{j-1})_{22}\nonumber\\
&&\quad(C_j)_{21}(A_{j+1}\cdots A_n)_{11}\nonumber\\
&=&
\frac{
(\frac{1}{2}\csch\xi e^\xi)^{i-1}(-k_c)^{j-i}
(\frac{1}{2}\csch\xi e^\xi)^{n-j}
}{(\frac{1}{2}\csch\xi e^\xi)^{n-1}}\nonumber\\
&=& e^{-\xi(j-i)},
\end{eqnarray}
such that the resulting $e^{-\xi(j-i)}$ can be re-factorized into
a product of factors $e^{-\xi}$ between $i$ and $j$. Therefore,
the bond dimension for representing $V_{ij}$ is reduced from 4 to 1,
which, when coupled with the operators $n_in_j$, leads to the MPO \eqref{expMPO} with $D=3$.

In 2D and 3D, such a simplification is unlikely to be possible,
thus our construction will lead to a TNO with bond dimension
$D=4D_O$. The factor 4 comes from the present TN construction for the
correlation functions $V_{ij}$, while $D_O$ depends on the
way that $V_{ij}$ is coupled with the product $n_in_j$
to form $\sum_{i<j}V_{ij}n_in_j$. It has been shown
that in 2D\cite{o2018efficient}, $D_O=3$ for the snake MPO construction
for $\sum_{i<j}n_in_j$, and $D_O=4$ using a 2D finite automata construction\cite{crosswhite2008finite,pirvu2010matrix,frowis2010tensor}.
In Sec. \ref{sec:3D}, we will show in 3D, $D_O$ can be 3, 4, or 5,
depending on whether an explicitly 1D, 2D, or 3D finite automata representation for $\sum_{i<j}n_in_j$ is used.

\section{2D formulation}\label{sec:2D}

\begin{figure}[ht]
\centering
\begin{tabular}{cc}
\includegraphics[width=0.2\textwidth]{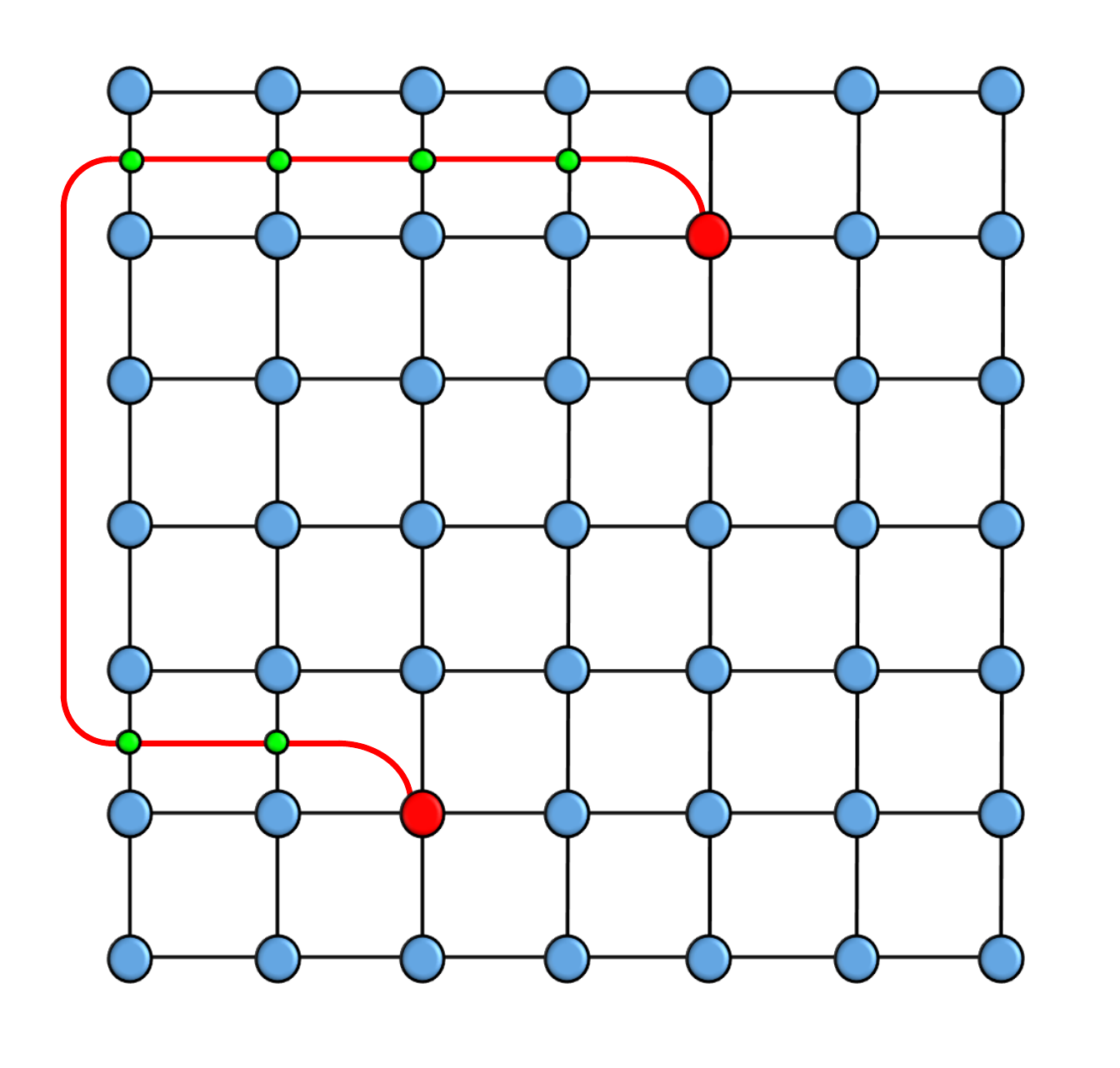} &
\includegraphics[width=0.2\textwidth]{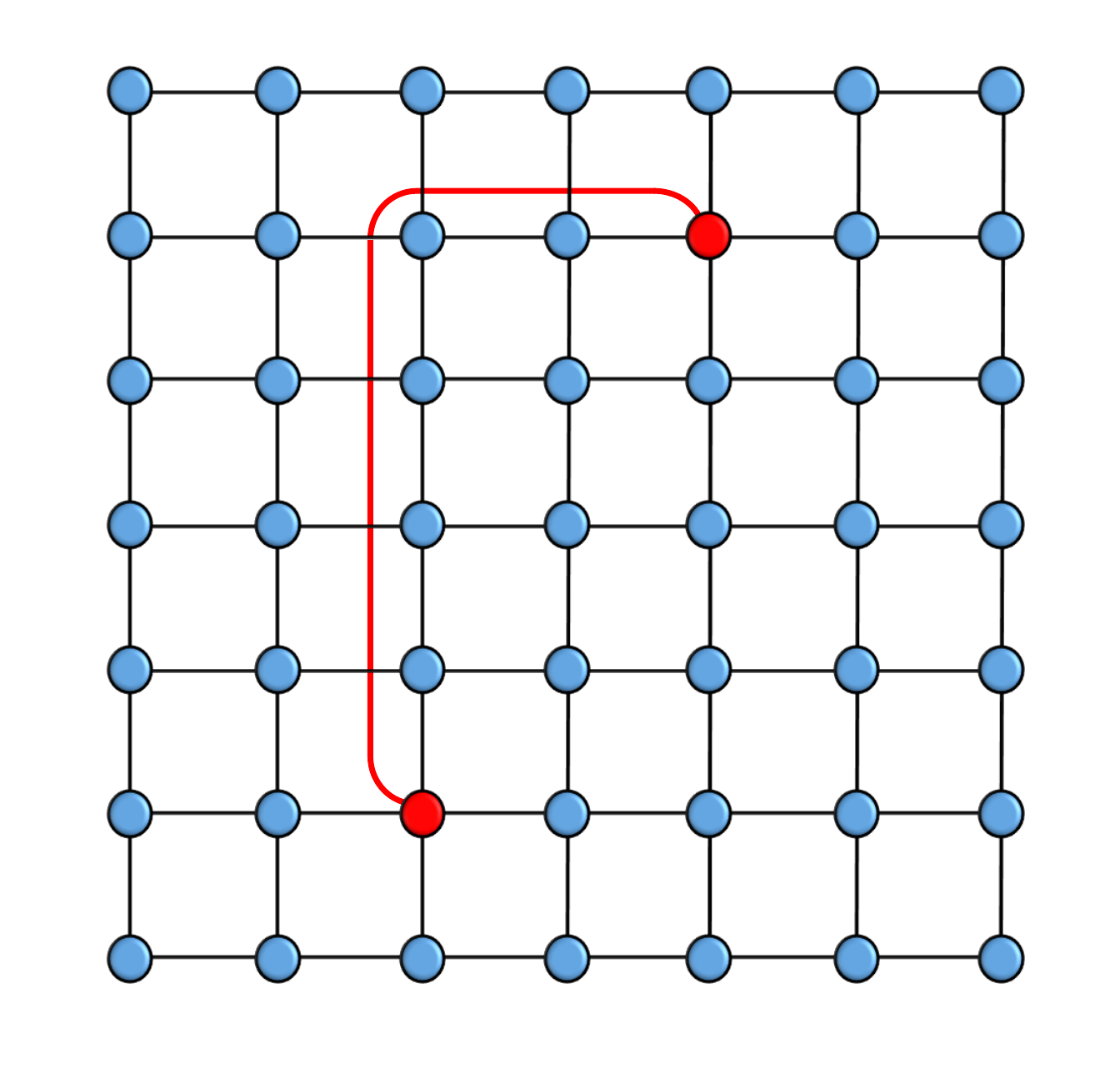} \\
(a) & (b) \\
\includegraphics[width=0.2\textwidth]{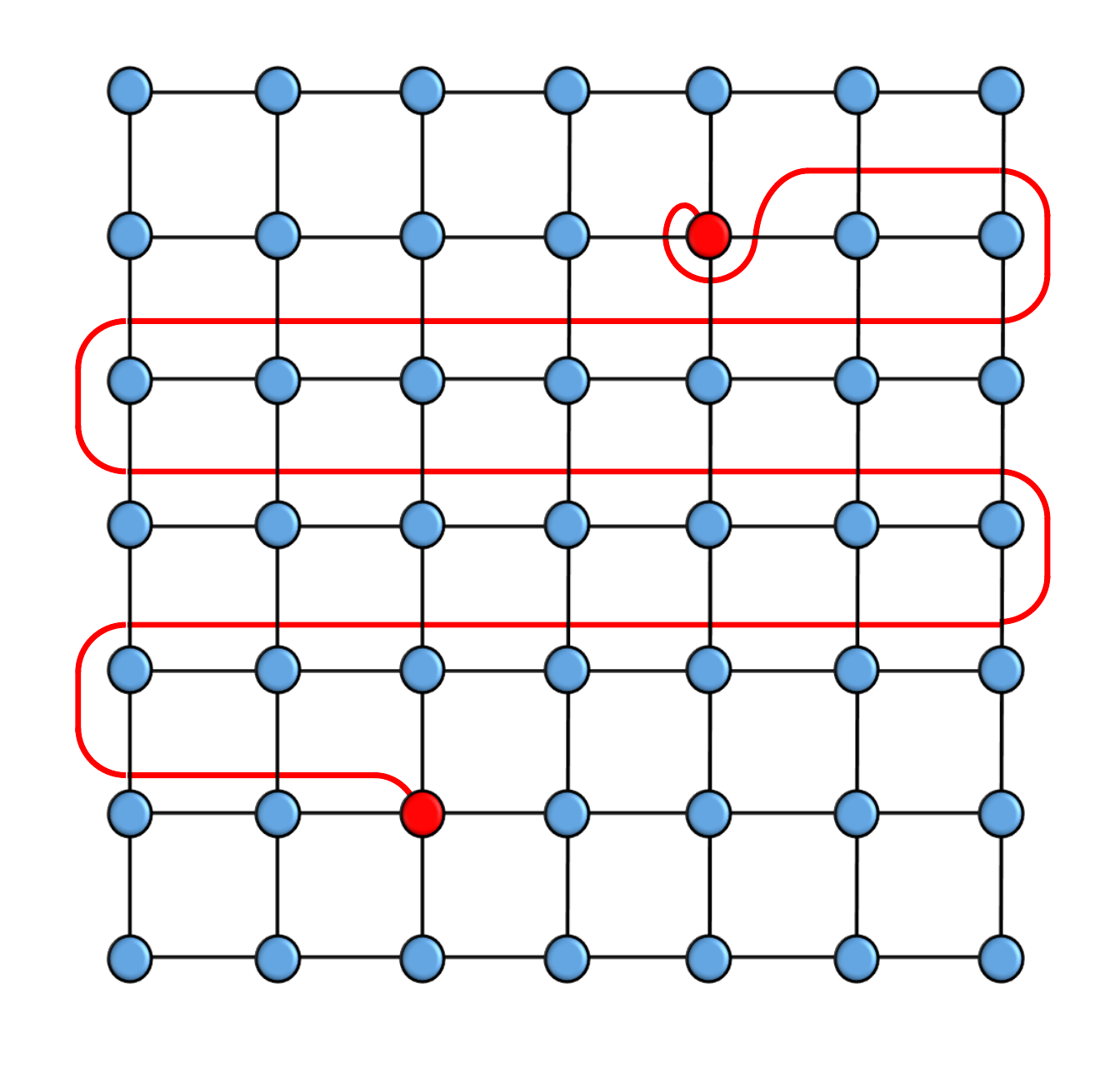} &
\includegraphics[width=0.2\textwidth]{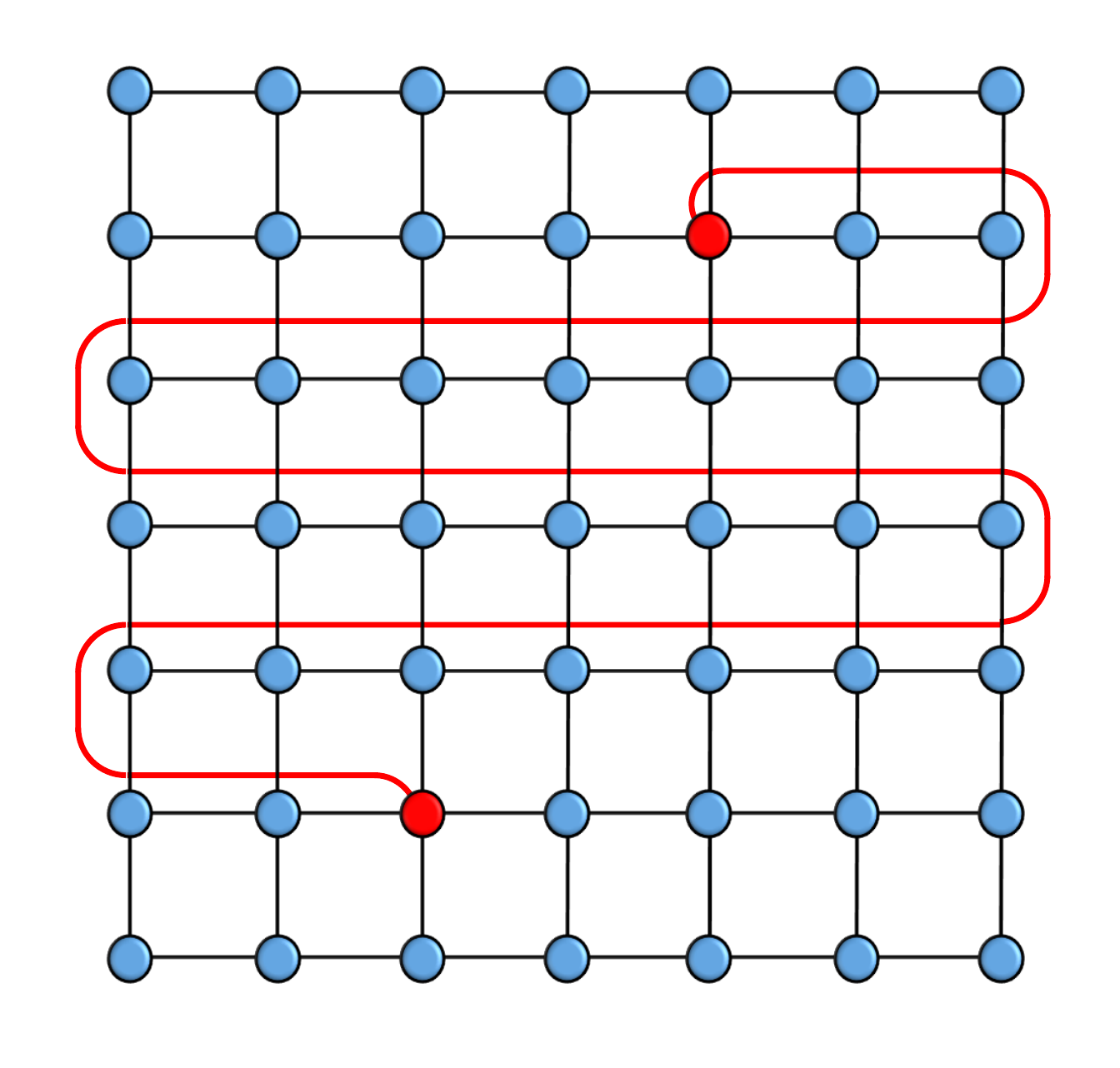} \\
(c) & (d) \\
\end{tabular}
\caption{Tensor network representation of correlation functions $\langle c_i\bar{c}_j\rangle$
in 2D: (a) basic representation with parity tensors (green dots) shown explicitly;
(b,c) fermionic paths are deformed from (a) and parity tensors (not shown explicitly for simplicity)
need to be inserted at each cross between the fermionic line (red) and the lattice.
TN representation (d) differs from (c) by -1 due to the jump over the site
containing $\bar{c}_j$.}\label{fig:rules2D}
\end{figure}

The problem of rewriting Eq. \eqref{Z3d} and the correlation functions
in 2D as products of local terms is more complicated than in 1D.
To avoid immediately delving into algebraic details,
we will first present the obtained results
in terms of graphical rules,
for which Figure \ref{fig:1d}(c) defines the local tensor configuration
and Figure \ref{fig:rules2D} defines the correlation functions.
Then a sketch of the derivation of these rules will be given
via a simple example.

\subsection{Rules for writing down TN representations}
By analogy to Eq. \eqref{localTensor1D} for 1D,
the local tensors for 2D are defined in the following way,
\begin{eqnarray}
(O_k)_{dlur} = \int \drm \bar{c}_k\drm c_k\; Q_k\delta_{k,d}\beta_{k,l}o_k\gamma_{k,u}\alpha_{k,r},\label{localTensor2D}
\end{eqnarray}
where $o_k$ can be one of $\{1,c_k,\bar{c}_k\}$ for $A_k$, $B_k$, or $C_k$, respectively.
While performing the integration manually as we did in 1D quickly becomes tedious for the different combinations
of subscripts $(d,l,u,r)$, the necessary integrals \eqref{localTensor2D} can be easily evaluated
using a simple program\cite{linkToLSnD}.
We can give Eq. \eqref{localTensor2D} a graphical representation as shown in
Figure \ref{fig:1d}(c), where the factors ($\delta,\beta,o,\gamma,\alpha$) appear
in a clockwise order starting from $\delta$. With this definition,
the 2D partition function $Z$ \eqref{Z3d} can be demonstrated to be given by
the PEPS $Z=\tr(\prod_k A_k)$. However, unlike in the 1D case,
the correlation function $V_{ij}$ is not simply
$\tr(\prod_{k\ne i,j} A_k B_i C_j)$.
Due to the anti-commutation of Grassmann variables,
some additional sign factors will appear when moving $c_i$ or
$\bar{c}_j$ to its local site. Assuming the site at position $(x,y)$ on an
$N$-by-$N$ lattice shown in Figure \ref{fig:rules2D}
is indexed by $(y-1)N+x$, one  finds that
there are additional sign factors such as $(-1)^{p(\gamma_k)}$ (or
equivalently $(-1)^{p(\delta_{k+N})}$) appearing at the position shown
by the green dot in Figure \ref{fig:rules2D}(a).
Here, $p(\gamma_k)$ is the parity of the bond $\gamma_k$ and is $0$ if $\gamma_k$ contains
an even number of Grassmanns and $1$ otherwise.
It is then seen that when representing $V_{ij}$ as a Grassmann correlation function
some parity tensors,
\begin{eqnarray}
(P_k)_{ab}=\delta_{ab}(-1)^{p(\gamma_{k,a})}=\mathrm{diag}(1,-1,-1,1),\label{parity}
\end{eqnarray}
need to be inserted on the upward bonds for the sites to the left of each fermionic site (red dot).
The final TN representation for $V_{ij}$ is given by Figure \ref{fig:rules2D}(a) (omitting the red lines).

From this basic representation, we can derive various equivalent
TN representations for $V_{ij}$ by using the parity conserving properties
of the tensor $O_k$ \eqref{localTensor2D}, which means that
if one of its virtual bonds is odd in parity, then the total parity of the other virtual bonds
must also be, $p(o_k)+1$ mod 2, otherwise the Grassmann integration vanishes.
This property allows us to define a jump move similar to that in fPEPS\cite{corboz2010simulation}.
Graphically, we can view the parity tensors as a result of the crossing
between a fermionic line (red) connecting sites $i$ and $j$ and the bonds between local tensors, see Figure \ref{fig:rules2D}(a). Then, starting
from this graph,
one is free to deform the fermionic line freely, as long as the necessary
parity factors are inserted at the crossings. This degree of freedom can be used
to make the fermionic line coincide with the path
used in the finite automata construction\cite{o2018efficient}
to derive rules for coupling with operators $n_in_j$, thus allowing
the parity factors to be inserted by the automata construction itself.
Figures \ref{fig:rules2D}(b) and \ref{fig:rules2D}(c)
are examples of a simple path and snake path, respectively.
This eventually allows for the construction of the
PEPO for $\sum_{i<j}V_{ij} n_in_j$, as shown in our previous work\cite{o2018efficient}.

Finally, it should be noted that Figure \ref{fig:rules2D}(c) and Figure \ref{fig:rules2D}(d) differ by a minus sign.
This is because when moving from (c) to (d), the jump through
$C_j$ \eqref{localTensor2D} introduces a minus sign
as $\bar{c}_j$ is odd, such that $(-1)^{p(\delta_{j,d})+p(\beta_{j,l})+p(\gamma_{j,u})+p(\alpha_{j,r})}
=-1$ for nonvanishing Grassmann
integrations. In summary, we can express both $Z$ and $V_{ij}$ in 2D
as PEPS, with the latter requiring additional parity factors on
certain virtual bonds given by Figure \ref{fig:rules2D}(a).

\subsection{Sketch of the derivations}

To illustrate how the above rules for 2D are actually derived, we consider a simple $4\times4$ example.
From Eq. \eqref{Z3d}, the partition function $Z$ can be rewritten as
\begin{eqnarray}
Z&=&
\int \Drm[\bar{c},c] (\prod_k Q_k)
[(\alpha_1\beta_2)\cdots]
[(\gamma_1\delta_5)\cdots]\nonumber\\
&=&
\int \Drm[\bar{c},c] \left(
[Q_1 \alpha_1(\gamma_1\delta_5)][Q_2\beta_2\alpha_2(\gamma_2\delta_6)]\cdots
\right),\label{Z2d}
\end{eqnarray}
where a more instructive way to write the right hand side is the following 2D representation:
{\small
\begin{eqnarray}
\begin{array}{llll}
\alpha_{13} & \beta_{14}\alpha_{14} &\beta_{15}\alpha_{15} & \beta_{16}\\
\alpha_{9}(\gamma_{9}\delta_{13}) & \beta_{10}\alpha_{10}(\gamma_{10}\delta_{14}) &
\beta_{11}\alpha_{11}(\gamma_{11}\delta_{15}) & \beta_{12}(\gamma_{12}\delta_{16})\\
\alpha_{5}(\gamma_{5}\delta_{9}) & \beta_{6}\alpha_{6}(\gamma_{6}\delta_{10}) &
\beta_{7}\alpha_{7}(\gamma_{7}\delta_{11}) & \beta_{8}(\gamma_{8}\delta_{12})\\
\alpha_1(\gamma_1\delta_5) & \beta_2\alpha_2(\gamma_2\delta_6) & \beta_3\alpha_3(\gamma_3\delta_7) &
\beta_4(\gamma_4\delta_{8})\\
\end{array}\label{2Dfactors}
\end{eqnarray}}with the even-parity factors $Q_k$ and the indices for components omitted for simplicity. In Eq. \eqref{2Dfactors}, one should read from the bottom-left factor $\alpha_1$ to the upper-right factor $\beta_{16}$ for $Z$ \eqref{Z2d}. In this representation, it is clear that in order to move all the factors to their local sites, we only need to
move all $\delta_k$ one row up in Eq. \eqref{2Dfactors} along the 1D sequence for $Z$. One way  we found to be convenient is to move them column-by-column from left to right.
That is, we first move $\delta_{13}$, $\delta_{9}$, and $\delta_{5}$ sequentially
to the respective upper rows, and then consider moving
$\delta_{14}$, $\delta_{10}$, and $\delta_{6}$, etc. We illustrate this
explicitly for $\delta_{13}$. When moving this past $\beta_{10}$ in the second column,
the factor $(-1)^{p(\delta_{13})p(\beta_{10})}$ appears
due to the exchange of $\delta_{13}$ with $\beta_{10}$.
Using the fact that the bond pairs \eqref{expDecomposition} are always even,
i.e., $p(\gamma_9\delta_{13})=1$ and $p(\alpha_9\beta_{10})=1$,
this factor can be made local $(-1)^{p(\delta_{13})p(\beta_{10})}=
(-1)^{p(\gamma_9)p(\alpha_9)}$, which can be further cancelled out
by a local exchange from $\alpha_9\gamma_9$ to $\gamma_9\alpha_9$ in the product \eqref{2Dfactors}.
This is how the ordering of factors in Figure \ref{fig:1d}(c) is derived.
After exchanging $\delta_{13}$ with $\beta_{10}$,
we move $\delta_{13}$ past $\alpha_{10}(\gamma_{10}\delta_{14})$, $\beta_{11}\alpha_{11}(\gamma_{11}\delta_{15})$,
$\beta_{12}(\gamma_{12}\delta_{16})$, but these can be regrouped into complete bonds, $(\alpha_{10} \beta_{11})(\gamma_{10}\delta_{14})\ldots$ which
are all even, thus no more signs accrue in moving $\delta_{13}$. Once the $\delta$ factors in the first column have been moved to their
local sites, these sites are in their final forms as shown in Eq. \eqref{Z2d}, where the product
of factors at each site is even and parity preserving.
Thus, when
 moving the factors in the second column, we can jump
over the sites in the first column without incurring any sign factor.
By repeating this procedure, we can express $Z$ as a PEPS
with local tensors defined in Figure \ref{fig:1d}(c).

The same process applies to the correlation functions. In this case,
taking $\langle c_7\bar{c}_{10}\rangle$ as an example,
the counterpart of Eq. \eqref{2Dfactors} is
{\small
\begin{eqnarray}
\begin{array}{llll}
\alpha_{13} & \beta_{14}\alpha_{14} &\beta_{15}\alpha_{15} & \beta_{16}\\
\alpha_{9}(\gamma_{9}\delta_{13}) & \beta_{10}\bar{c}_{10}\alpha_{10}(\gamma_{10}\delta_{14}) &
\beta_{11}\alpha_{11}(\gamma_{11}\delta_{15}) & \beta_{12}(\gamma_{12}\delta_{16})\\
\alpha_{5}(\gamma_{5}\delta_{9}) & \beta_{6}\alpha_{6}(\gamma_{6}\delta_{10}) &
\beta_{7}c_7\alpha_{7}(\gamma_{7}\delta_{11}) & \beta_{8}(\gamma_{8}\delta_{12})\\
\alpha_1(\gamma_1\delta_5) & \beta_2\alpha_2(\gamma_2\delta_6) & \beta_3\alpha_3(\gamma_3\delta_7) &
\beta_4(\gamma_4\delta_{8})\label{2DfactorsVij}\\
\end{array}
\end{eqnarray}}The task is again to move the $\delta$ factors to their local sites,
and we can apply the same procedure for Eq. \eqref{2DfactorsVij}.
However, one can see that moving $\delta_{13}$ will involve an additional
exchange with $\bar{c}_{10}$,
which results in an additional sign factor $(-1)^{p(\delta_{13})}$.
A similar situation occurs when moving $\delta_{9}$ and $\delta_{10}$
due to the exchanges with $c_{7}$. These additional sign factors
give the rule for parity tensors $(P_k)_{ab}$ (green dots) in Figure \ref{fig:rules2D}(a).
Thus, the long-range interaction $V_{ij}=\langle c_i\bar{c}_j\rangle$ is given by the quotient
of the TN diagram in Figure \ref{fig:rules2D}(a) and that for $Z$.

\section{3D formulation}\label{sec:3D}
\subsection{Rules for writing down TN representations}
\begin{figure}[ht]
\centering
\begin{tabular}{c}
\includegraphics[width=0.3\textwidth]{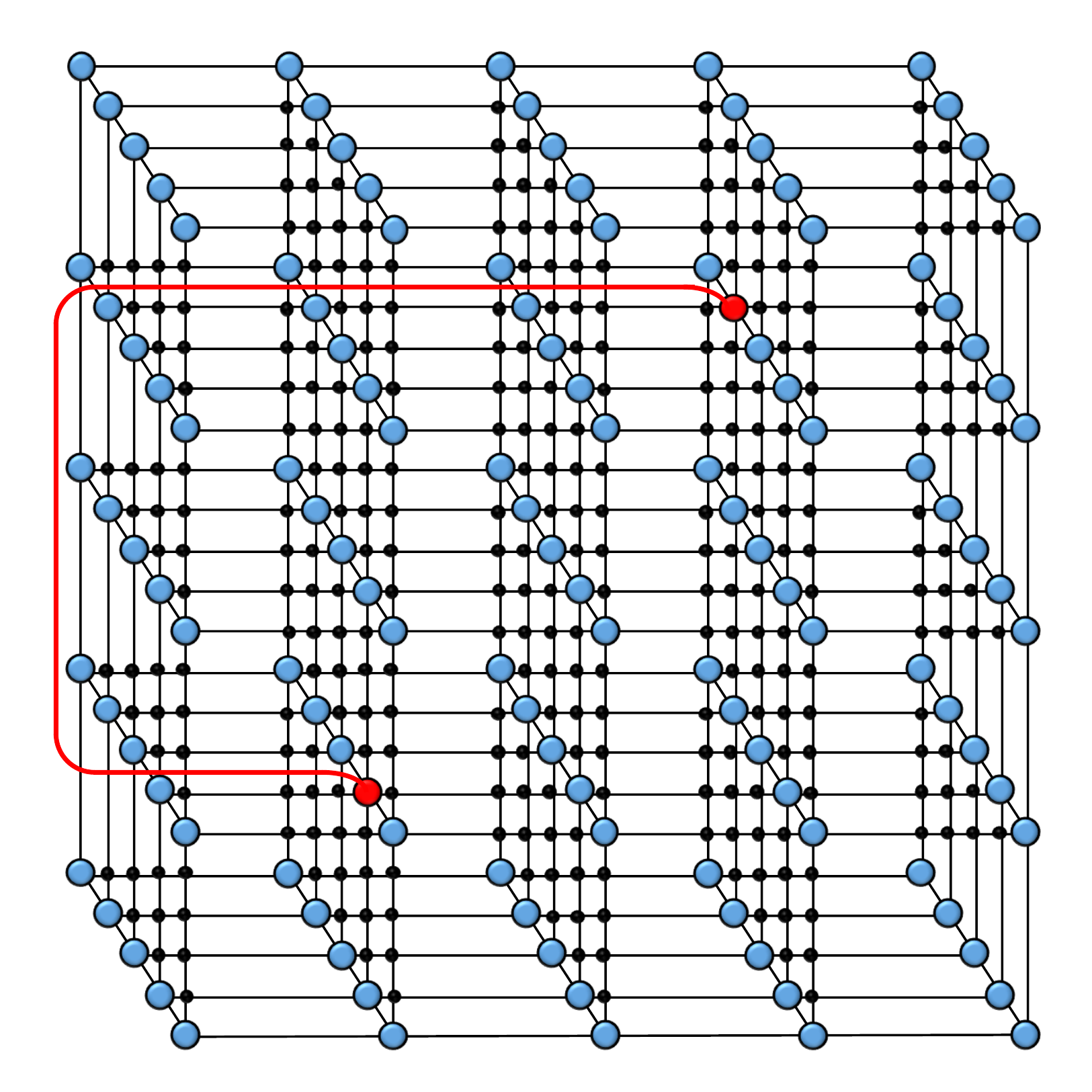} \\
(a) Tensor network (TN) representation for $V_{ij}$ in 3D \\
\includegraphics[width=0.3\textwidth]{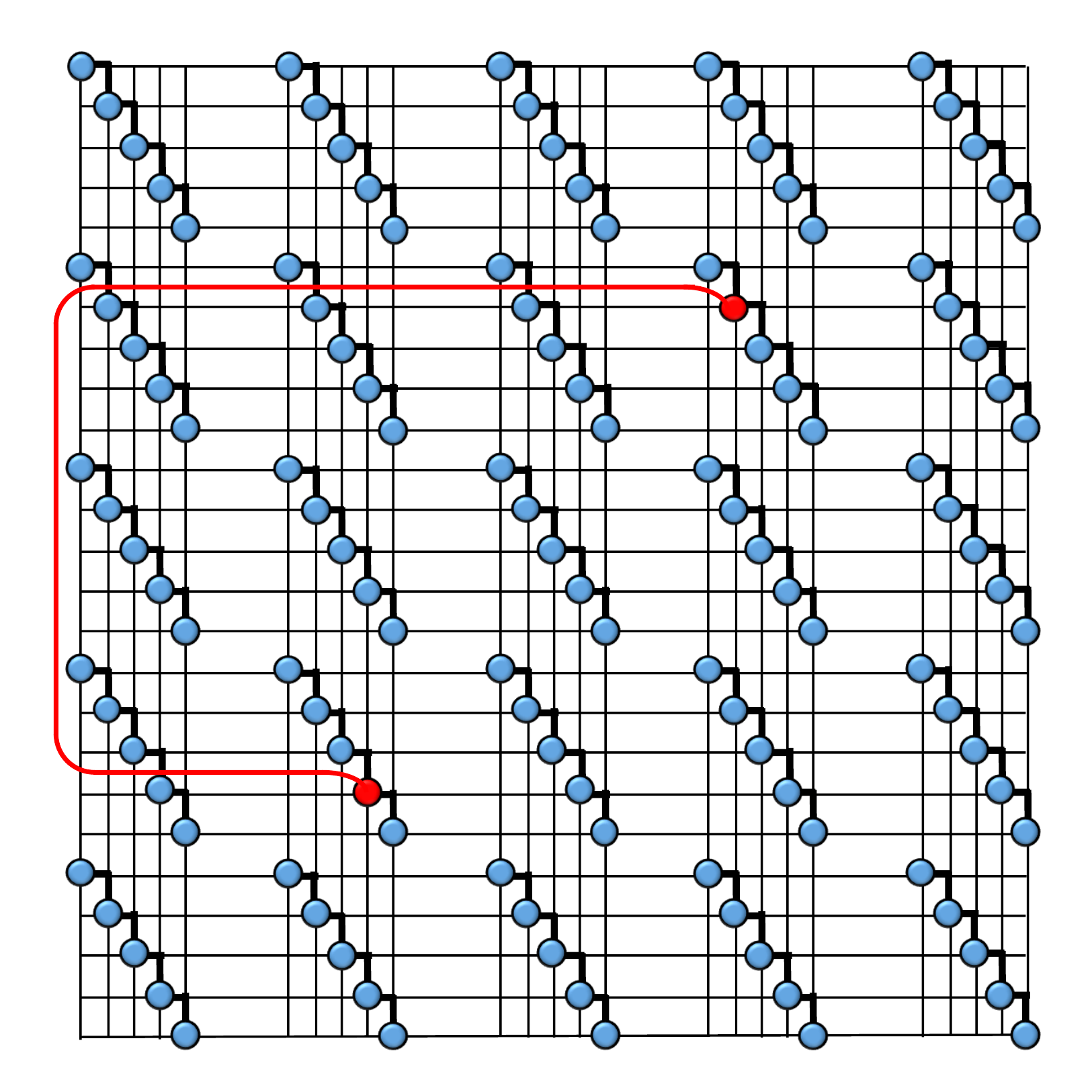} \\
(b) One way to contract 3D TN as PEPS \\
\end{tabular}
\caption{Tensor network representation of correlation functions $\langle c_i\bar{c}_j\rangle$
in 3D: (a) 3D TN representation for $V_{ij}$,
(b) An $N\times N\times N$ 3D network can be mapped to an $N^2\times N^2$ 2D network with $N=5$
for contractions using algorithms for PEPS, where the diagonal bonds between physical sites
have been folded into the square lattice as highlighted by
the bold black lines.}\label{fig:rules3D}
\end{figure}

Similar to the 2D graphical representation, the local tensors in 3D
can be written down according to Figure \ref{fig:1d}(c), viz.,
\begin{eqnarray}
(O_k)_{dblutr} = \int \drm \bar{c}_k\drm c_k\; Q_k
\delta_{k,d}\tau_{k,b}\beta_{k,l}o_k\gamma_{k,u}\sigma_{k,t}\alpha_{k,r},\label{localTensor3D}
\end{eqnarray}
where $o_k\in\{1,c_k,\bar{c}_k\}$ for $A_k$, $B_k$, or $C_k$, respectively,
and where the Grassmann integral can be conveniently evaluated via the same program\cite{linkToLSnD}.
However, the partition function $Z$ in 3D \eqref{Z3d}
is not simply given by $\tr(\prod_k A_k)$ as in 1D and 2D.
The correct TN representation is given in Figure
\ref{fig:rules3D}(a), where a \emph{swap} tensor (black dot, see also
Figure \ref{fig:swap}),
\begin{eqnarray}
S^{wx}_{yz}=\delta_{wz}\delta_{xy}(-1)^{p(w)p(x)},\label{swap}
\end{eqnarray}
needs to be introduced at each crossing between
a vertical bond and a horizontal bond, when
the 3D network is viewed as a projection onto 2D.
The necessity for these swap tensors is explained in Sec.
\ref{sec:3Dswap}. [NB: The special case of an $N\times N\times 2$ 3D network is structurally identical to the network for the overlap $\langle\Psi|\Psi\rangle$ between two fPEPS\cite{corboz2010simulation}.]
For correlation functions, the rule
for the TN representation can still be summarized by the fermionic line (red) in Figure \ref{fig:rules3D}(a).
We will discuss the derivation of this rule given by Figure \ref{fig:sign3D} in the next section.

Before closing this section for the 3D rules, we mention that
the resulting $N\times N\times N$ 3D network can also be viewed
 equivalently as a $N^2\times N^2$ 2D network, see
Figure \ref{fig:rules3D}(b), which can be readily contracted using
standard algorithms for 2D PEPS. This mapping also implies that
to construct the 3D TNO for $\sum_{i<j}V_{ij}n_in_j$,
we can use the same finite automata rules used in 2D\cite{o2018efficient} (either the explicitly 2D rules with $D_O=4$ or
the 1D snake MPO rules with $D_O=3$) to construct the tensor network representation
of the operator sum $\sum_{i<j}n_in_j$.
This can be seen by indexing the physical site at the position $(x,y,z)$ by
$(z-1)N^2+(y-1)N+x$, such that the relative ordering of physical sites
is unchanged when mapped into 2D. In addition to
these rules for the operators, one can also use a set of ``3D'' rules
with $D_O=5$ to construct the TNO representation of $\sum_{i<j}n_in_j$,
which explicitly uses the 3D lattice structure, see Appendix.
Thus, the final 3D TNO representation for $\sum_{i<j}V_{ij}n_in_j$
will have bond dimension $D=4D_O$, where $D_O$ can be chosen to be 3, 4, or 5.

\subsection{Sketch of the derivations}
While the above rules for expressing $V_{ij}$ as a TN may look familiar to readers who have previously worked with fPEPS,
in this section, for a more general audience, we will give a pedagogical explanation of
two of the main ingredients in the derivations: (1) in the TN representation of $Z$
in 3D \eqref{Z3d} (Figure \ref{fig:rules3D}(a)),
how we obtain the order of factors in Eq. \eqref{localTensor3D} for the local tensors, and how the swap tensors arise,
(2) in the TN representation of $V_{ij}$, how the rule for the fermionic line (red) is derived.

\subsubsection{Partition function, local tensors, and swap tensors}\label{sec:3Dswap}

\begin{figure}[ht]
\centering
\begin{tabular}{cc}
\includegraphics[width=0.2\textwidth]{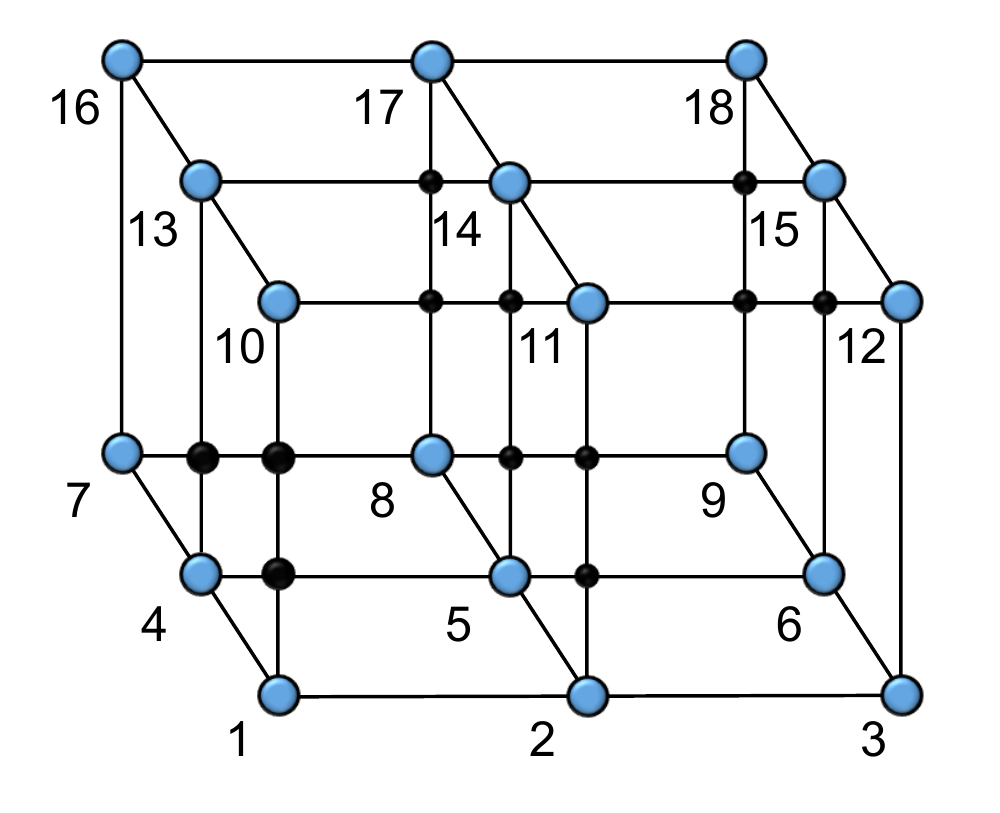} &
\includegraphics[width=0.15\textwidth]{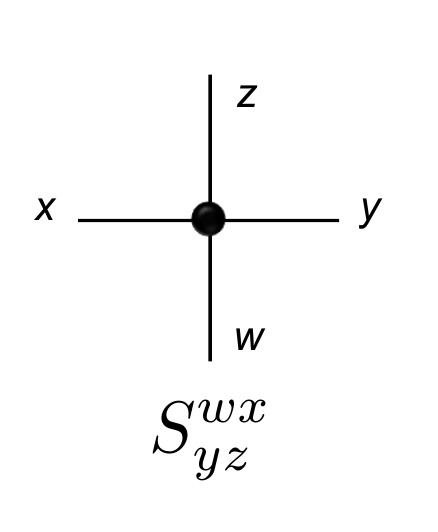} \\
\end{tabular}
\caption{Example for the partition function $Z$ of a simple $3\times3\times2$ lattice, including the swap tensors (black dots).
  The swap tensors near the bottom-left corner tensors, which are
  elaborated on in Sec. \ref{sec:3Dswap}, are represented by larger black dots.}\label{fig:swap}
\end{figure}

For simplicity, we consider a simple $3\times3\times2$ lattice shown in Figure \ref{fig:swap}.
From Eq. \eqref{Z3d}, the partition function $Z$ can be rewritten as
\begin{eqnarray}
Z&=&
\int \Drm[\bar{c},c] (\prod_k Q_k)
[(\alpha_1\beta_2)\cdots]
[(\gamma_1\delta_4)\cdots]
[(\sigma_1\tau_{10})\cdots]\nonumber\\
&=&
\int \Drm[\bar{c},c] \left(
[Q_1 \alpha_1(\gamma_1(\sigma_1\tau_{10})\delta_4)\beta_2]
\cdots\right),\label{Z3d332}
\end{eqnarray}
where the integrand can be written simply as {\small
\begin{eqnarray}
\begin{array}{lll}
\alpha_{16} & \beta_{17}\alpha_{17} & \beta_{18} \\
\alpha_{13}(\gamma_{13}\delta_{16}) & \beta_{14}\alpha_{14}(\gamma_{14}\delta_{17}) & \beta_{15}(\gamma_{15}\delta_{18}) \\
\alpha_{10}(\gamma_{10}\delta_{13}) & \beta_{11}\alpha_{11}(\gamma_{11}\delta_{14}) & \beta_{12}(\gamma_{12}\delta_{15}) \\
\alpha_7(\sigma_7\tau_{16}) & \beta_8\alpha_8(\sigma_8\tau_{17}) & \beta_9(\sigma_9\tau_{18}) \\
\alpha_4(\gamma_4(\sigma_4\tau_{13})\delta_7) & \beta_5\alpha_5(\gamma_5(\sigma_5\tau_{14})\delta_8) & \beta_6(\gamma_6(\sigma_6\tau_{15})\delta_9) \\
\alpha_{1}(\gamma_{1}(\sigma_{1}\tau_{10})\delta_{4}) & \beta_{2}\alpha_2(\gamma_2(\sigma_2\tau_{11})\delta_{5})
& \beta_{3}(\gamma_3(\sigma_3\tau_{12})\delta_{6})
\end{array}\label{3Dfactors}
\end{eqnarray}}which, similarly to Eq. \eqref{2Dfactors}, should
be read from bottom-left to upper-right. Now to move
all factors to local sites, apart from the need to move $\delta$ up one row
as in the 2D case, the $\tau$ factors also need to be moved up one layer, which
increases the complexity of finding the TN representation of $Z$ in 3D.

Similarly to in the 2D case, we found the most convenient way to move factors in 3D  to be face-by-face from left to right.
The $\delta$ factors ($\delta_4$, $\delta_7$, $\delta_{13}$, $\delta_{16}$) can first be moved
to their respective local site in the same way as in 2D, viz., {\small
\begin{eqnarray}
\begin{array}{lll}
\delta_{16}\alpha_{16} & \beta_{17}\alpha_{17} & \beta_{18} \\
\delta_{13}\gamma_{13}\alpha_{13} & \beta_{14}\alpha_{14}(\gamma_{14}\delta_{17}) & \beta_{15}(\gamma_{15}\delta_{18}) \\
\gamma_{10}\alpha_{10} & \beta_{11}\alpha_{11}(\gamma_{11}\delta_{14}) & \beta_{12}(\gamma_{12}\delta_{15}) \\
\delta_7\alpha_7(\sigma_7\tau_{16}) & \beta_8\alpha_8(\sigma_8\tau_{17}) & \beta_9(\sigma_9\tau_{18}) \\
\delta_{4}\gamma_4\alpha_4(\sigma_4\tau_{13}) & \beta_5\alpha_5(\gamma_5(\sigma_5\tau_{14})\delta_8) & \beta_6(\gamma_6(\sigma_6\tau_{15})\delta_9) \\
\gamma_{1}\alpha_{1}(\sigma_{1}\tau_{10}) & \beta_{2}\alpha_2(\gamma_2(\sigma_2\tau_{11})\delta_{5})
& \beta_{3}(\gamma_3(\sigma_3\tau_{12})\delta_{6})
\end{array}
\end{eqnarray}}where we have exchanged the $\gamma$ and $\alpha$ factors in the first column to compensate
for the introduced sign factors. Next, we move the $\tau$ factors in the order $\tau_{16}$, $\tau_{13}$, and $\tau_{10}$,
which is essential for simplifying the manipulations, to the upper layer, leading to {\small\begin{eqnarray}
\begin{array}{lll}
\tau_{16}\delta_{16}\alpha_{16} & \beta_{17}\alpha_{17f} & \beta_{18} \\
\tau_{13}\delta_{13}\gamma_{13}\alpha_{13} & \beta_{14}\alpha_{14}(\gamma_{14}\delta_{17}) & \beta_{15}(\gamma_{15}\delta_{18}) \\
\tau_{10}\gamma_{10}\alpha_{10} & \beta_{11}\alpha_{11}(\gamma_{11}\delta_{14}) & \beta_{12}(\gamma_{12}\delta_{15}) \\
\delta_7\alpha_7\sigma_7 & \beta_8\alpha_8(\sigma_8\tau_{17}) & \beta_9(\sigma_9\tau_{18}) \\
\delta_{4}\gamma_4\alpha_4\sigma_4 & \beta_5\alpha_5(\gamma_5(\sigma_5\tau_{14})\delta_8) & \beta_6(\gamma_6(\sigma_6\tau_{15})\delta_9) \\
\gamma_{1}\alpha_{1}\sigma_{1} & \beta_{2}\alpha_2(\gamma_2(\sigma_2\tau_{11})\delta_{5})
& \beta_{3}(\gamma_3(\sigma_3\tau_{12})\delta_{6})
\end{array}
\end{eqnarray}}Note that after moving $\tau_{16}$ to the upper row, the factors in site 7 are complete,
such that when moving $\tau_{13}$, no additional sign factors due to the jump over this site need to be considered.
Thus, the net sign factors introduced are $(-1)^{p(\tau_{16})p(\beta_8)}$,
$(-1)^{p(\tau_{13})[p(\beta_5)+p(\beta_8)]}$, and $(-1)^{p(\tau_{10})[p(\beta_2)+p(\beta_5)+p(\beta_8)]}$, respectively.
Again by noting the even parity of bonds, the factors such as $(-1)^{p(\tau_{16})p(\beta_8)}$ can be made local
$(-1)^{p(\tau_{16})p(\beta_8)}=(-1)^{p(\sigma_7)p(\alpha_7)}$, and further absorbed locally by exchanging
$\alpha_7$ and $\sigma_7$ in the local product $\delta_7\alpha_7\sigma_7$. These local exchanges
to compensate the local sign factors determine the order of factors in
Eq. \eqref{localTensor3D} or equivalently Figure \ref{fig:1d}(c). The remaining
signs that cannot be absorbed are given by $(-1)^{p(\tau_{13})p(\beta_8)}(-1)^{p(\tau_{10})p(\beta_5)}
(-1)^{p(\tau_{10})p(\beta_8)}$. These nonlocal terms can be exactly represented/decomposed
in terms of swap tensors (black dots) shown in Figure \ref{fig:swap}.
The whole process for moving $\delta$ and $\tau$ factors can be repeated for the other faces/columns such that
the final TN representation of $Z$ is given by a 3D network composed of local tensors $A_i$ and swap tensors.

\subsubsection{Correlation functions and fermionic line}
\begin{figure}[ht]
\centering
\begin{tabular}{cc}
\includegraphics[width=0.22\textwidth]{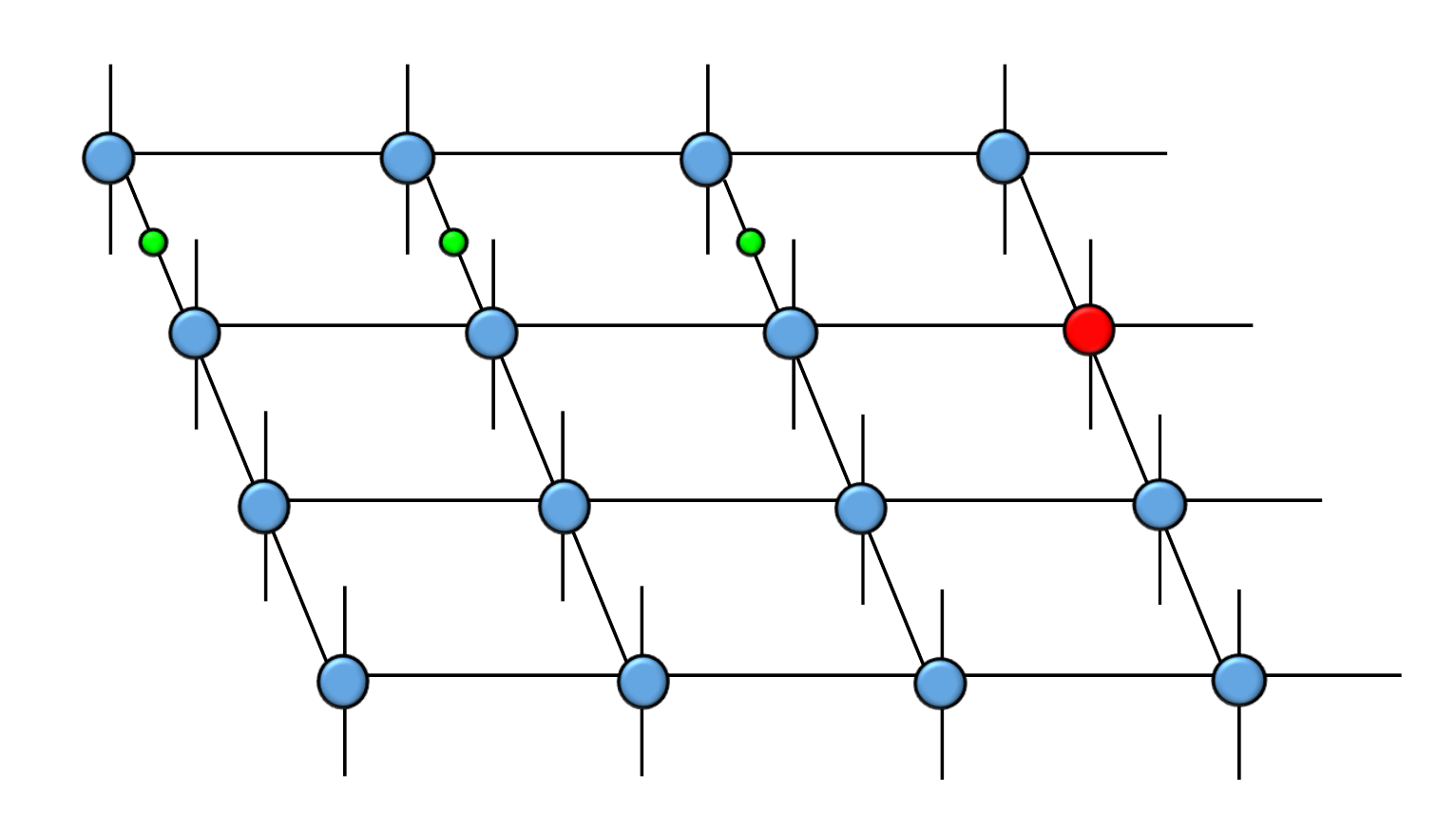} &
\includegraphics[width=0.22\textwidth]{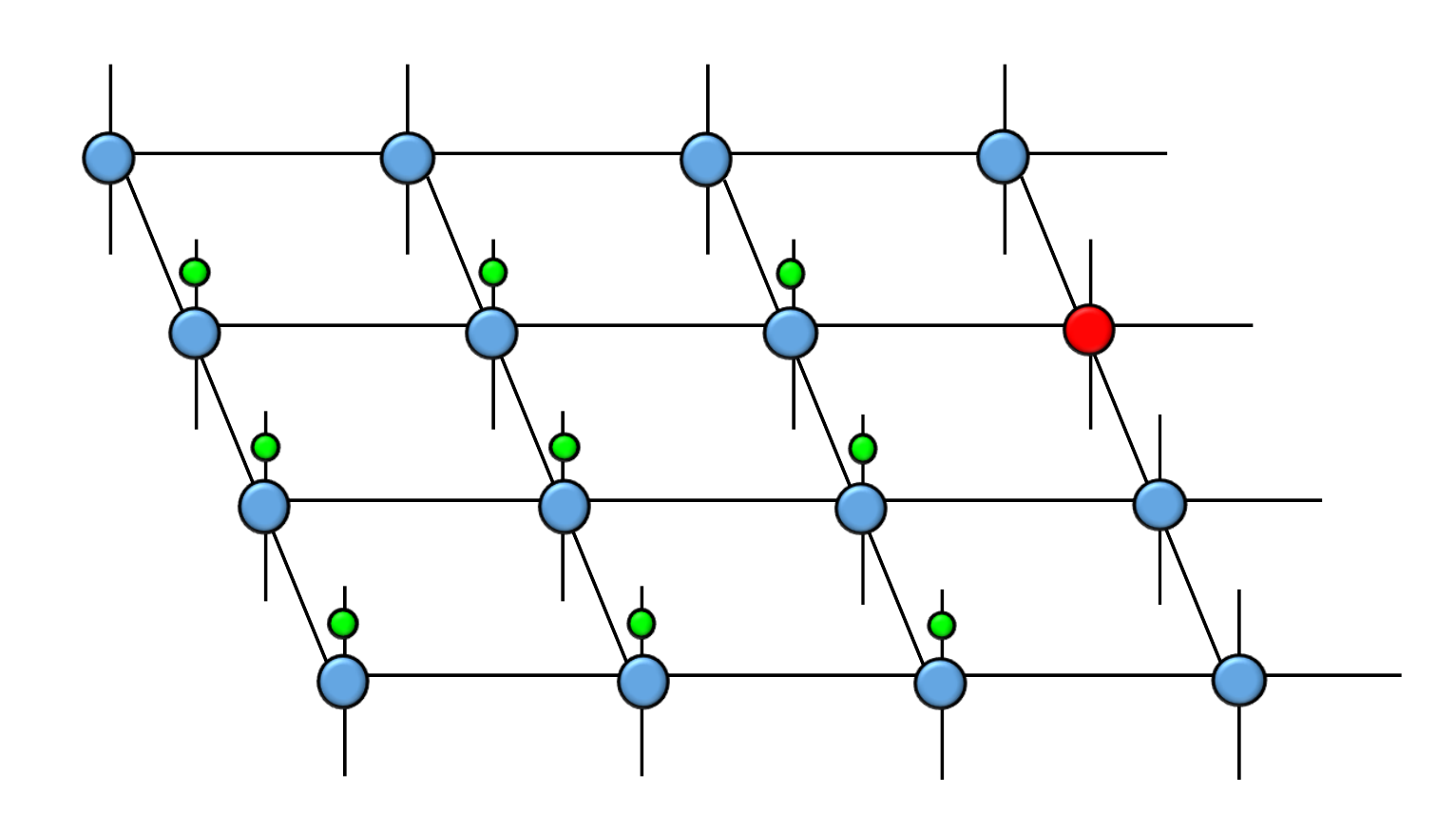} \\
(a) & (b) \\
\includegraphics[width=0.22\textwidth]{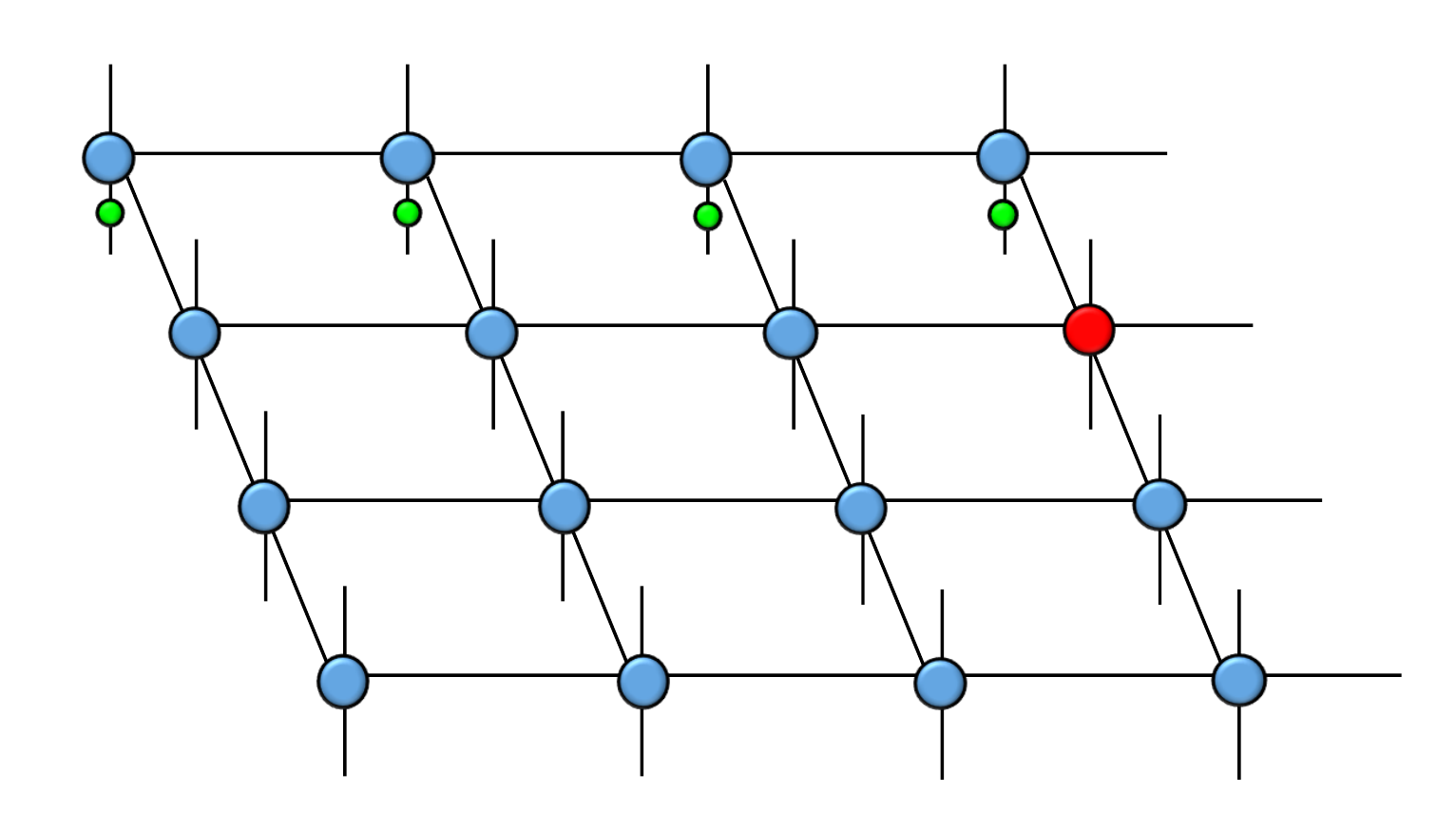} &
\includegraphics[width=0.22\textwidth]{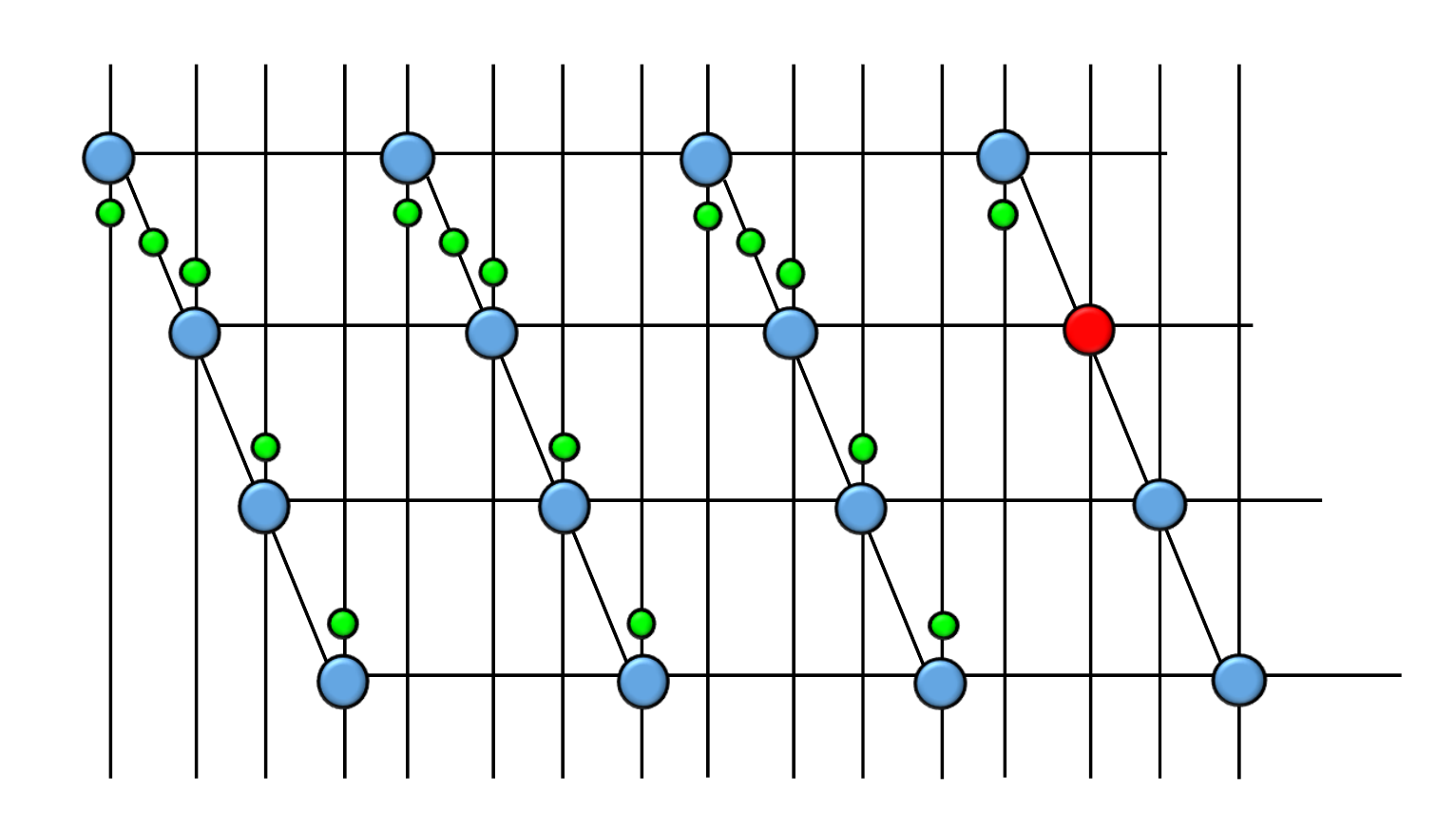} \\
(c) & (d) \\
\includegraphics[width=0.22\textwidth]{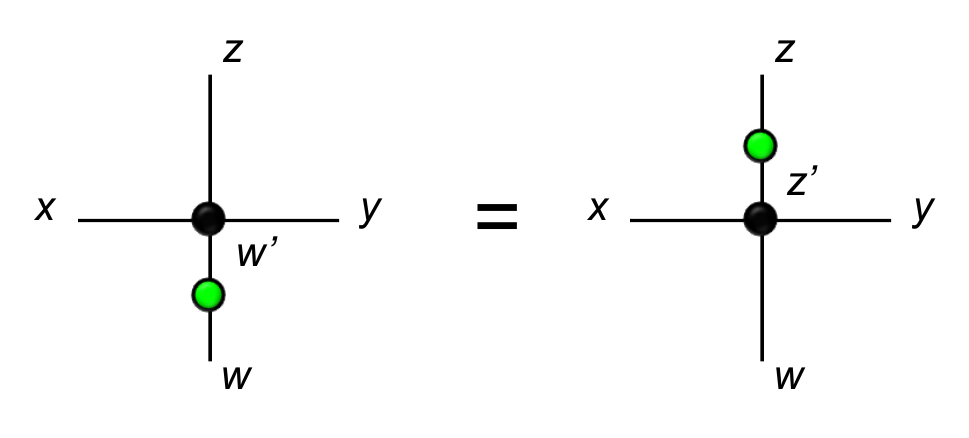} &
\includegraphics[width=0.22\textwidth]{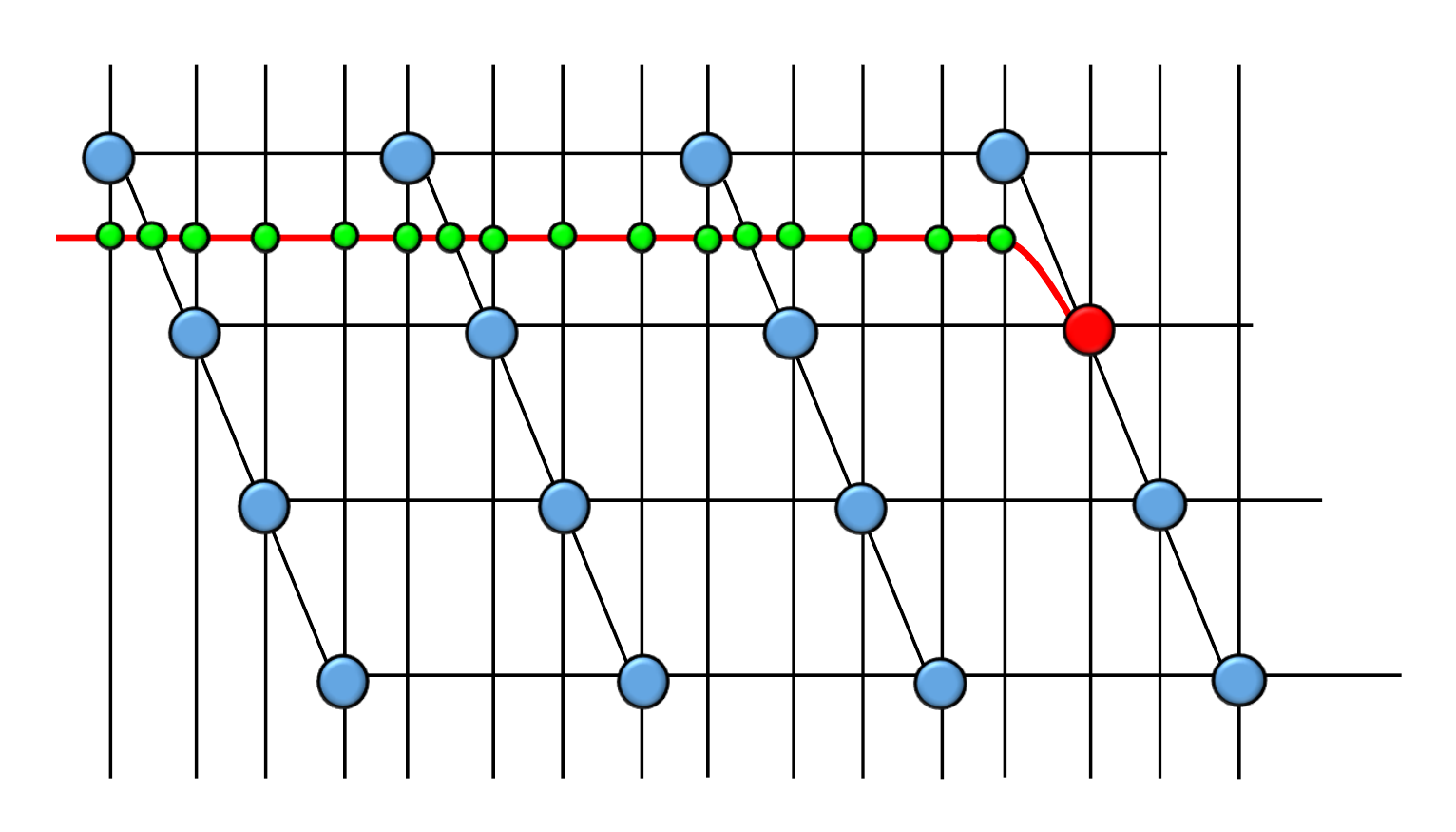} \\
(e) & (f) \\
\end{tabular}
\caption{Illustration of the derivation of rules for representing correlation functions in 3D:
(a,b,c) three sets of parity tensors appear in the 3D case due to the odd parity of $c_i$ (or $\bar{c}_{j}$);
(d) all parity tensors  together with swap tensors on the line intersections (not explicitly shown for simplicity);
(e) the exchange of swap and parity tensors; (f) the transformation of (d) into an equivalent
rule for fermionic crossing by moving certain parity tensors upwards using the exchange rule (e).}\label{fig:sign3D}
\end{figure}

For the correlation functions $\langle c_i\bar{c}_j\rangle$ in 3D, the introduced parity factors
can be found in the same manner following the logic for 2D. Thus, we only describe
the basic idea here, assuming $c_i\bar{c}_j$ (even parity and $i<j$) is first placed on  site $i$,
which means $\bar{c}_j$ needs to be moved to site $j$.
One can show that the additional parity factors introduced by the fermionic variables
$c_i$ ($\bar{c}_j$) can be classified into three groups for $c_i$ ($\bar{c}_j$),
as shown in Figures \ref{fig:sign3D}(a,b,c), for crossings with different bonds.
Specifically, the in-plane parity factors for $c_i$ ($\bar{c}_j$) in Figure \ref{fig:sign3D}(a) are the same as
those in 2D, see Figure \ref{fig:rules2D}(a), while Figures \ref{fig:sign3D}(b,c) are
new due to the existence of bonds in the $z$-direction. Summarizing
all parities and swaps together leads to Figure \ref{fig:sign3D}(d),
which can be greatly simplified into a single rule
of a fermionic line (red) in Figure \ref{fig:sign3D}(f)
by moving certain parities upwards using the exchange rule shown in Figure \ref{fig:sign3D}(e),
viz., $\sum_{w'}P_{ww'}S^{w'x}_{yz}=(-1)^{p(w)}S^{wx}_{yz}=(-1)^{p(z)}S^{wx}_{yz}=\sum_{z'}S^{wx}_{yz'}P_{z'z}$
following from the definitions in \eqref{parity} and \eqref{swap}.
Therefore, the final rule shown in Figure \ref{fig:sign3D}(f) for half of the fermionic pair
and Figure \ref{fig:rules3D}(a) for the whole pair $c_i\bar{c}_j$
is the same as that for 2D, see Figure \ref{fig:rules2D}(a).

\section{Conclusions}\label{sec:Conclusion}
In this work we have presented an analytic construction of
the tensor network representations of the discretized
Green's function $V_{ij}$ of the Helmholtz equation 
in 2D and 3D using Grassmann Gaussian integration.
The resulting TN representation is very compact,
with bond dimension $D=4$. Interestingly, in 3D it gives
an analytic TN representation of correlation functions decaying
as $r_{ij}^{-1}$ asymptotically, which yields
the discretized (screened) Coulomb interaction on the simple cubic lattice.
The TN representation can be made compatible with
the rules for finite automata by properly deforming the associated fermionic lines,
such that we can construct a TNO representation for $\sum_{i<j}V_{ij}n_in_j$. 
These interactions with different Helmholtz parameters can be used as basis
functions to fit decaying long-range interactions in higher dimensions,
as an analog of the exponential fitting procedure used in TN algorithms in 1D.

The resulting TN operators can readily be used in simulations of
continuum systems, such as the uniform electron gas (UEG),
via the combination of a discretized lattice representation and tensor network algorithms. Another possible
direction is the simulation of the effective low-energy sectors of lattice gauge theory\cite{banuls2018tensor}
using higher dimensional TNS. Integrating out gauge
degrees of freedom leads to Hamiltonians with non-local or long-range terms.
In fact, this is precisely how the Coulomb interaction in Nature arises.

\section*{Acknowledgements}
This work was supported by the US National Science Foundation via Grant No. 1665333. ZL is supported by the Simons Collaboration on the Many-Electron Problem. MJO is supported by a US National Science Foundation Graduate Research Fellowship under Grant No. DEG-1745301. GKC is a Simons Investigator in Physics.

\section*{Appendix: Finite automata rules for coupling with operators in 3D}\label{appendix}
In this section, we discuss how to construct a 3D PEPO representation for the
distance-independent interaction operator,
\begin{equation}
\hat{V}=\sum_{i < j} \hat{A}_i \hat{B}_j,
\label{eqn:AiBj}
\end{equation}
with a bond dimension $D_O=5$. In other words, we will build $\hat{V}$ as
$\hat{V}=\tr(\prod_k\hat{P}_k)$ with $(\hat{P}_k)_{L_k,U_k,D_k,R_k,T_k,B_k}=O_{n_kn_k'}^{[k]}$
being a local operator on site $k$. This PEPO representation can be combined\cite{o2018efficient}
with the 3D PEPS representation for
correlation functions $V_{ij}$ described before (Sec. \ref{sec:3D}), resulting in a PEPO representation of
$\sum_{i < j} V_{ij}\hat{A}_i \hat{B}_j$ with $D=4D_O=20$.

\begin{figure}[ht]
\begin{center}
\begin{tabular}{c}
\includegraphics[width=0.3\textwidth]{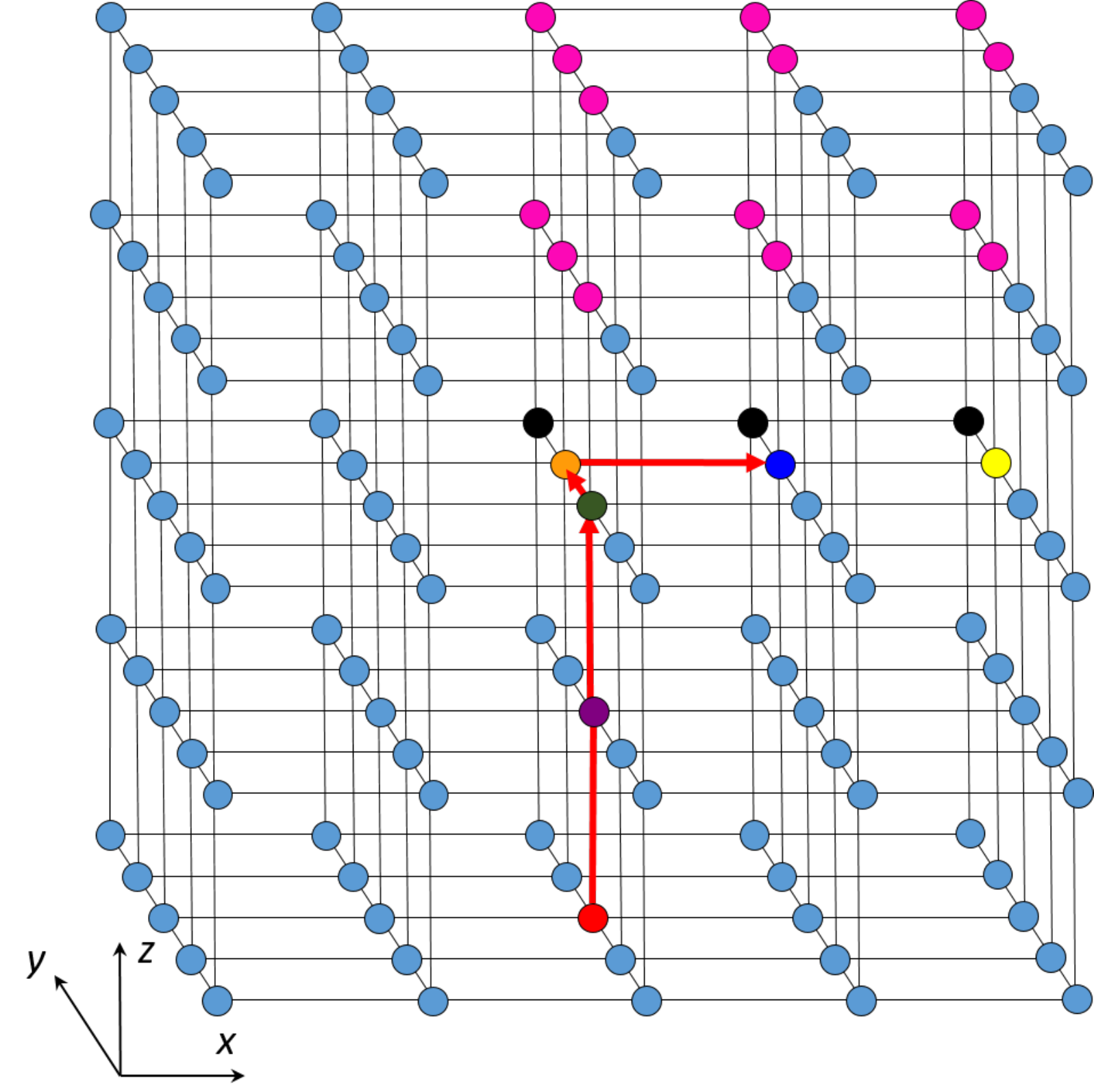}\\
(a) \\
\includegraphics[width=0.3\textwidth]{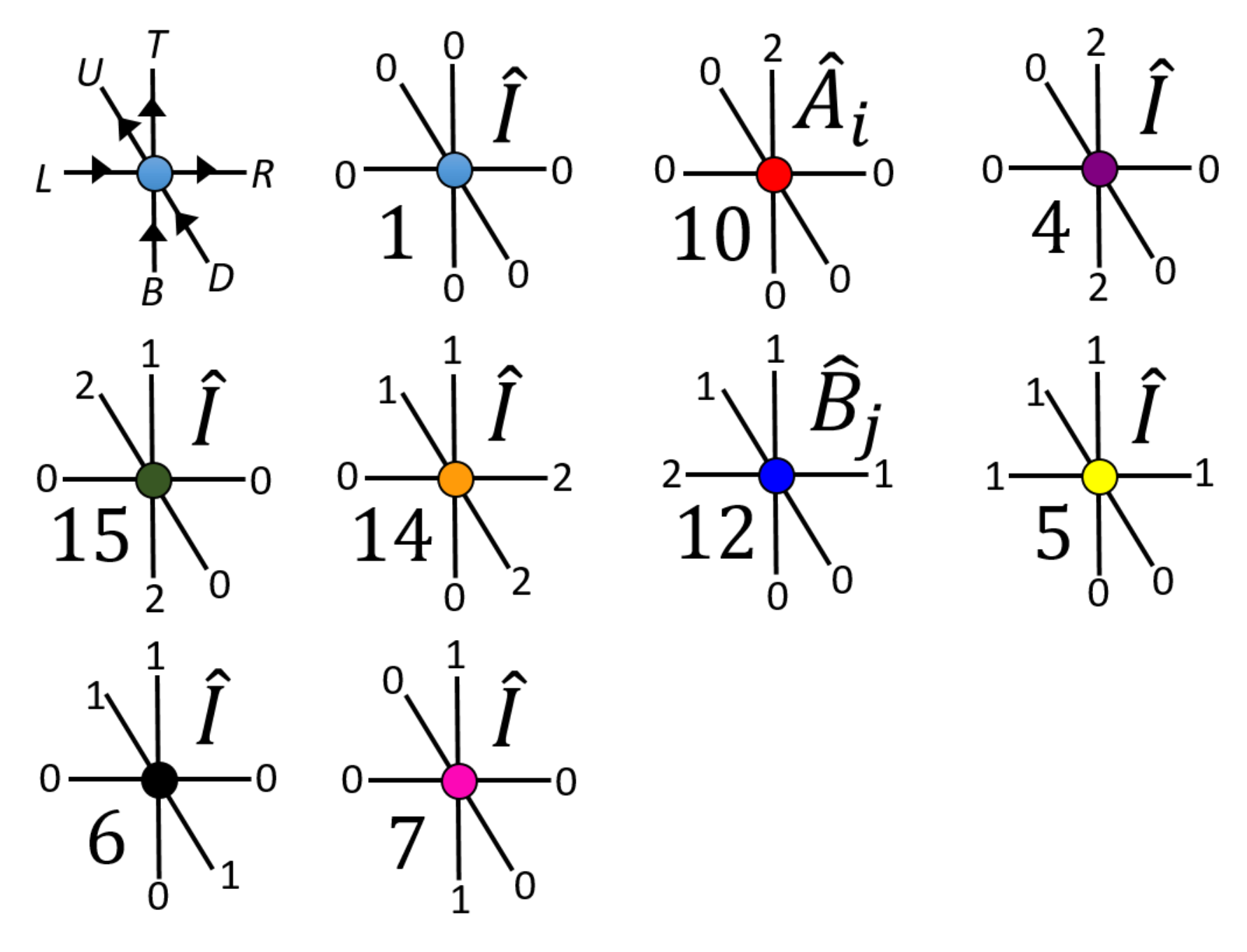}\\
(b) \\
\end{tabular}
\caption{This figure and those that follow are examples of the set of
rules needed to construct the operator-valued
3D finite automata that encodes the pairwise interaction PEPO $\sum_{i<j} \hat{A}_i \hat{B}_j$
for arbitrary operators $\hat{A}$ and $\hat{B}$. The color of the tensor
in (a) corresponds to the index configuration of the equivalently colored
tensor in (b). In (b), the local operator corresponding to the given index configuration
is given to the top-right of each tensor, and the rule number of each tensor is given to
its bottom-left. For tensors along the boundary, the relevant legs are simply
removed from the corresponding diagram in (b). This specific case shows
an interaction between $\hat{A}_{i}$ (red) and $\hat{B}_{j}$ (dark blue),
where the signal path between the two sites is shown in red.}
\label{fig:3drules_xpypzp}
\end{center}
\end{figure}

The finite automata (also known as finite state machine) picture\cite{crosswhite2008finite,pirvu2010matrix,frowis2010tensor}
of a PEPO views each tensor as a node in a graph, and
each virtual bond of dimension $D_O$ as a directed
edge in the graph that can pass $D_O$ different signals
(or has $D_O$ different possible states). By convention
we have chosen our directed edges to point in the
$+\hat{x}$, $+\hat{y}$, and $+\hat{z}$ directions, where
the axes are defined in Figure \ref{fig:3drules_xpypzp}. This allows us to impose
an ordering of the sites, where we traverse the $x$ direction most quickly, then
$y$, then $z$, starting from the bottom left corner.
With this convention, the tensor at position $k$ has its
$U_k$, $R_k$, and $T_k$ indices (corresponding to $+\hat{y}$, $+\hat{x}$, and $+\hat{z}$ respectively) pass ``outgoing'' signals while
its $L_k$, $D_k$ and $B_k$ indices (corresponding to $-\hat{x}$, $-\hat{y}$, and $-\hat{z}$)
receive ``incoming'' signals (see the first tensor in Fig. \ref{fig:3drules_xpypzp}(b)).
When a tensor has certain specific combinations of incoming
and outgoing signals, that tensor's two physical indices $n_k$ and $n'_k$ (not shown
in Fig. \ref{fig:3drules_xpypzp} for simplicity) encode
a local operator $O^{[k]}_{n_k n'_k}$, which is either a physical
operator ($\hat{A}$, $\hat{B}$) or the identity operator ($I$).
These special combinations
of index values precisely correspond to the set of rules which generate a
finite state machine (PEPO) which encodes
all the terms in the sum \eqref{eqn:AiBj}.
In order to avoid encoding any additional unwanted terms,
the value of $O^{[k]}_{n_k n'_k}$ is the zero operator $\hat{0}_k$
when the states of the six virtual indices do not match any rule which generates
the desired machine. In other words, unwanted
configurations of the state machine (and thus unwanted
configurations of the local operators) are prevented by causing such a configuration
to trigger the action
of $\hat{0}_k$ on at least one site of the machine, rendering the entire term null.

The complete list of the rules that define the full
3D PEPO which generates all pairwise interactions
in Eq.~\eqref{eqn:AiBj} with bond dimension
$D_O=5$ is given in Table \ref{tab:3drules}. The presentation
is in the style of Ref.~\cite{frowis2010tensor}. In the following,
we will provide an intuitive explanation for the derivation of these
rules, assuming some familiarity with the simpler constructions
in 1D and 2D\cite{crosswhite2008finite,frowis2010tensor,o2018efficient}.
 The present 3D construction can be viewed as a generalization
of the 2D rules by incorporating an additional set of rules
to include interactions between sites in different layers.

\begin{table}[ht]
\begin{center}
\begin{tabular}{c|c|c}
\hline\hline
	Rule number &
\begin{tabular}{c}
Index values \\ $(L_k,U_k,D_k,R_k,T_k,B_k)$
\end{tabular}
    & $O^{[k]}_{n_k n'_k}$ \\
	\hline
	1 & (0,0,0,0,0,0)  & $I_k$ \\
	2 & (0,2,2,0,1,0)  & $I_k$ \\
	3 & (2,1,0,2,1,0)  & $I_k$ \\
	4 & (0,0,0,0,2,2)  & $I_k$ \\
	5 & (1,1,0,1,1,0)  & $I_k$ \\
	6 & (0,1,1,0,1,0)  & $I_k$ \\
	7 & (0,0,0,0,1,1)  & $I_k$ \\
    \hline
    8 & (0,2,0,0,0,0)  & $\hat{A}_k$ \\
    9 & (0,1,0,2,0,0)  & $\hat{A}_k$ \\
    10 & (0,0,0,0,2,0) & $\hat{A}_k$ \\	
    11 & (0,1,2,1,1,0) & $\hat{B}_k$ \\
    12 & (2,1,0,1,1,0) & $\hat{B}_k$ \\
    13 & (0,1,0,1,1,2) & $\hat{B}_k$ \\
    \hline
    14 & (0,1,2,2,1,0) & $I_k$ \\
    15 & (0,2,0,0,1,2) & $I_k$ \\
    16 & (0,1,0,2,1,2) & $I_k$ \\
    \hline
    17 & (3,1,0,3,1,0) & $I_k$ \\
    18 & (3,1,2,1,1,0) & $I_k$ \\
    19 & (0,1,0,3,1,0) & $\hat{B}_k$ \\
    20 & (3,1,0,1,1,2) & $I_k$ \\
    \hline
    21 & (0,1,4,0,1,2) & $I_k$ \\
    22 & (0,4,4,0,1,0) & $I_k$ \\
    23 & (3,4,0,1,1,0) & $I_k$ \\
    24 & (0,4,0,2,1,0) & $I_k$ \\
    25 & (0,4,0,1,1,0) & $\hat{B}_k$ \\
    \hline
    $26^*$ & $P^{\text{upper right top corner}}_{0,0,0,0,0,0}$  & $\hat{0}_k$ \\
\hline\hline
\end{tabular}
\end{center}
\caption{The rules for the full 3D PEPO that generates all pairwise
interactions in Eq.~\eqref{eqn:AiBj} with $D_O=5$.
All combinations of indices not listed in this table correspond to
$O^{[k]}_{n_k n'_k}=\hat{0}_k$, while $I_k$ is
simply the identity operator. Note that the local operators
$\hat{A}_k$ and $\hat{B}_k$ do not have to be the same,
although in our case from the main text they would both be the number operator $n_k$.}
\label{tab:3drules}
\end{table}

\subsection{Basics}
A useful way to reason about the construction of finite state machines
is to assign some verbal meaning to each of the $D_O$ possible signals
that can be passed between the nodes. In the present case we have $D_O=5$, meaning
that each virtual bond of a tensor can take index values of (0, 1, 2, 3, and 4).
The meanings that we assign to these signals are used to describe the different ``messages''
of information that they pass to the adjacent tensor that the bond is connected to.
The ``0'' signal is the default signal, which generally means that the machine is in
its initial state along that signal path and no physical operators have been applied yet.
``1'' is the ``stop'' signal which, when received, generally tells a tensor to avoid acting
with a physical operator but instead to act with the identity operator. This is used when
another tensor along that signal path has applied a physical operator and
does not want an interaction to be generated along the direction that it sends the ``1'' message.
``2'' is the ``start'' signal, which is passed along the directed edges starting
with the action of $\hat{A}$ on site $i$ and terminating with the action of $\hat{B}$ on site $j$.
The path of this signal can be thought of as the ``interaction path.''

With these signals, we can encode all the terms in Eq.~\eqref{eqn:AiBj} for which
$j$ lies in the $+\hat{x}$/$+\hat{y}$/$+\hat{z}$ direction (or along the edges of this sector, for which $\Delta x$, $\Delta y$, or $\Delta z$ can be zero) with respect to site $i$. The rules
which generate these terms are 1-16. Rules 1-7 encode the propagation of the ``0'',
``1'', and ``2'' signals along the directed edges in straight lines. Rules 8-13 encode the
action of the physical operators, which begin and terminate ``1'' and ``2'' signals.
Rules 14-16 allow
for the ``2'' signal to ``turn'' in allowed directions. Specifically, 14 allows a ``2'' which is
travelling in the $+\hat{y}$ direction (and is thus received by the $D_k$ index) to turn and
propagate along the $+\hat{x}$ direction. Rule 15 encodes a turn from $+\hat{z}$ to $+\hat{y}$
and 16 allows a turn from $+\hat{z}$ to $+\hat{x}$. Note that other turns which do not violate the
directions of the edges, such as $+\hat{x}$ to $+\hat{y}$ and $+\hat{y}$ to $+\hat{z}$, are not allowed in order to prevent double counting. This illuminates a more subtle convention that we have chosen: for interactions
$\hat{A}_i \hat{B}_j$ in which sites $i$ and $j$ do not lie along a straight line, the ``2'' signal
first propagates in the $+\hat{z}$ direction, then $+\hat{y}$, then $+\hat{x}$ (when $i$ and $j$
are in the same plane but not along a straight line, one of the directions in this ordering is skipped). Figure~\ref{fig:3drules_xpypzp} provides a characteristic example of a set of tensor
configurations which encodes one ``basic'' interaction term.

\subsection{Remaining terms}
There are additional terms in the sum~\eqref{eqn:AiBj} for which site $j$ does not lie in the
$+\hat{x}$/$+\hat{y}$/$+\hat{z}$ direction with respect to site $i$. Specifically,
there are six additional cases:
\begin{enumerate}
\item $\{ -\hat{x}, +\hat{y}, \Delta \hat{z} = 0 \}$,
\item $\{ -\hat{x}, \Delta \hat{y} = 0,  +\hat{z} \}$,
\item $\{ -\hat{x}, +\hat{y}, +\hat{z} \}$,
\item $\{ \Delta \hat{x} = 0, -\hat{y}, +\hat{z} \}$,
\item $\{ +\hat{x}=0, -\hat{y}, +\hat{z} \}$,
\item $\{ -\hat{x}, -\hat{y}, +\hat{z} \}$.
\end{enumerate}
Since $j$ lies in at least one negative direction with respect to $i$, the ``2'' signal
that starts at site $i$ cannot propagate all the way to $j$ because at some point
it will need to go against the direction of a directed edge. To account for these
terms, the ``3'' and ``4'' signals can be introduced to propagate from site $j$ towards
site $i$ along the $+\hat{x}$ and $+\hat{y}$ directions, respectively. To complete the
interaction, these new signals
can then meet up with the ``2'' that began propagating from site $i$ towards site $j$ along
the $+\hat{x}$/$+\hat{y}$/$+\hat{z}$ directions via the introduction of new
state machine rules. Below we will explain case-by-case how this is done.



\begin{figure}[ht]
\begin{center}
\begin{tabular}{c}
\includegraphics[width=0.3\textwidth]{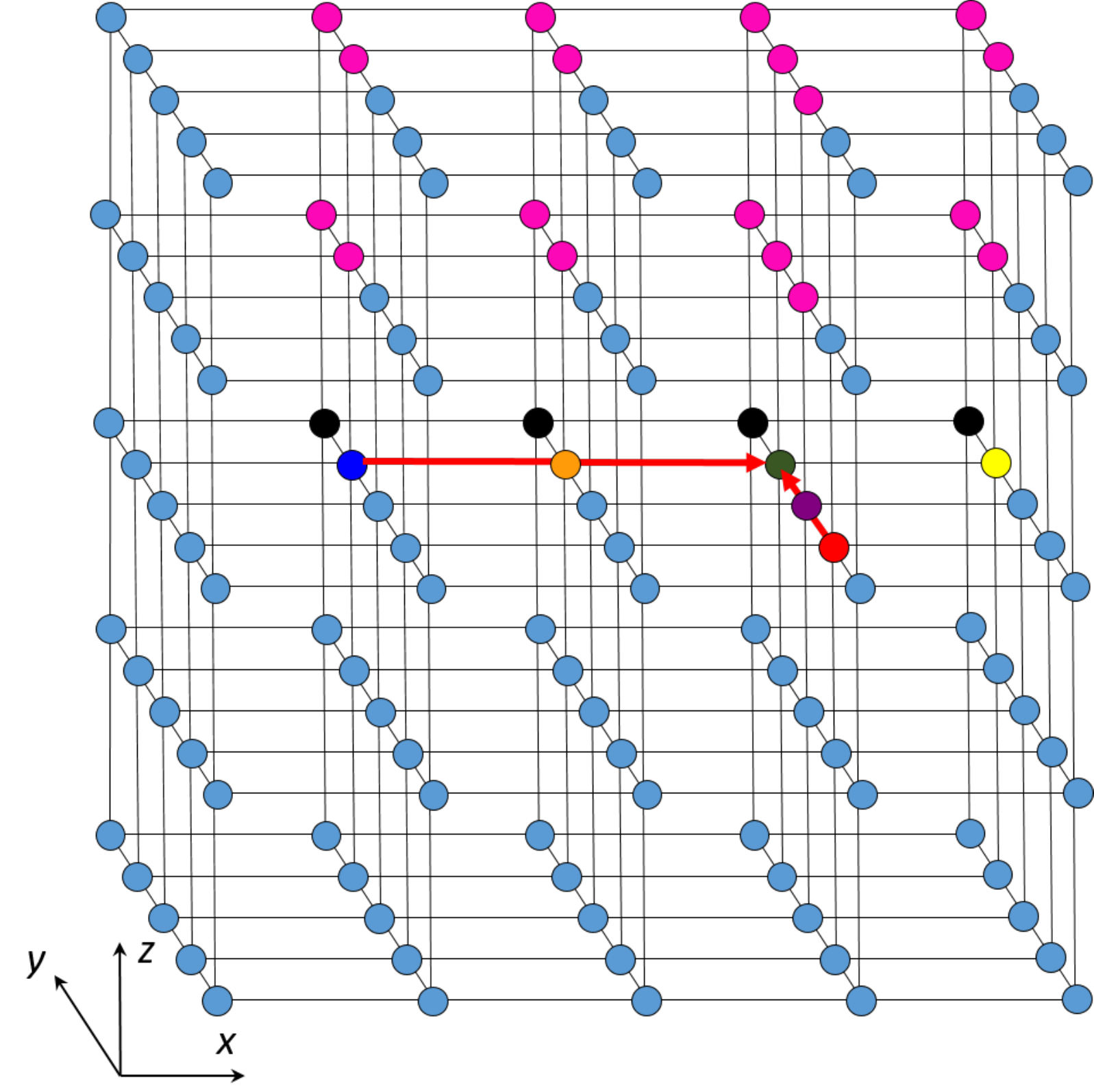}\\
(a) \\
\includegraphics[width=0.3\textwidth]{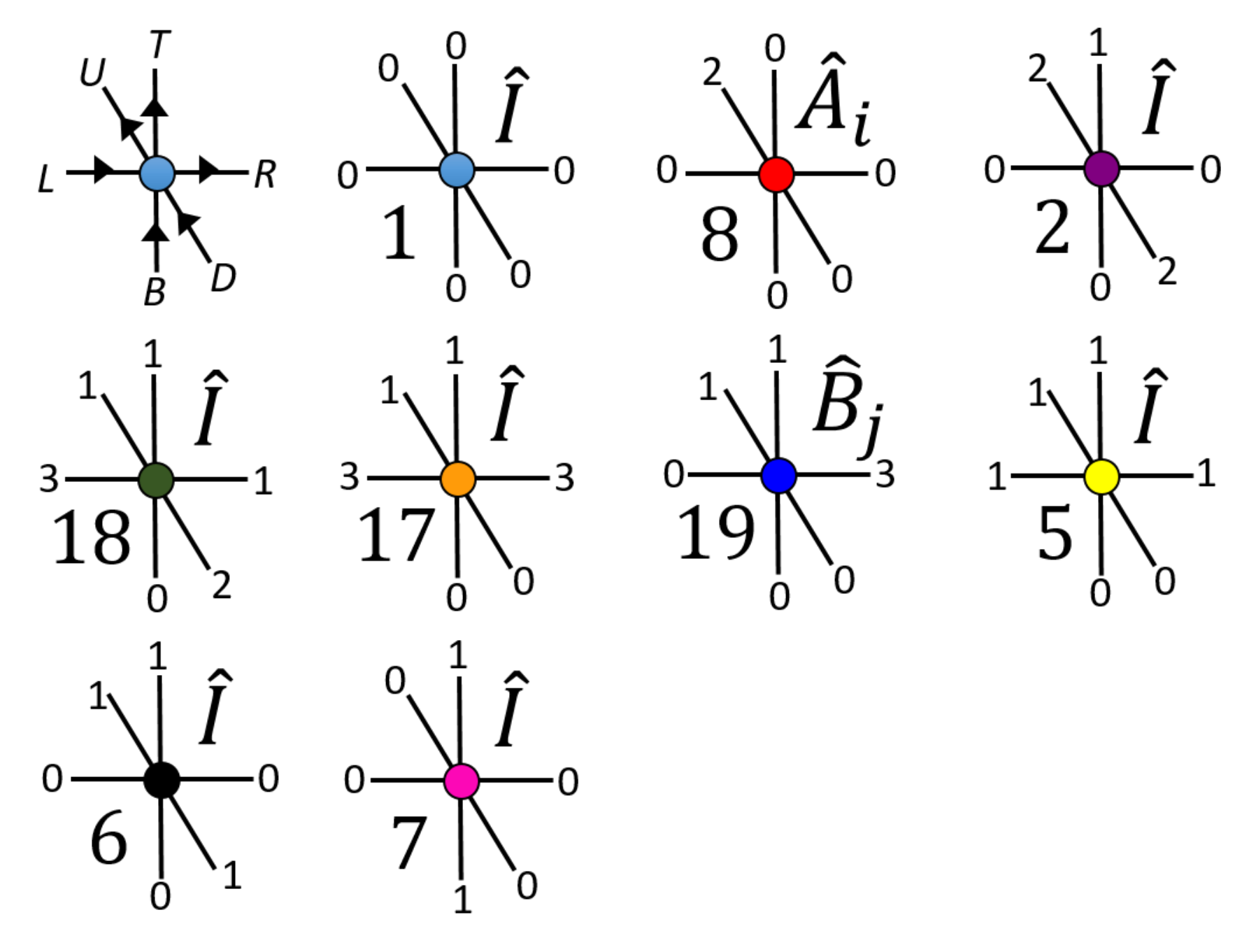}\\
(b) \\
\end{tabular}
\caption{Case 1: $\{ -\hat{x}, +\hat{y}, \Delta \hat{z} = 0 \}$.}
\label{fig:3drules_xmypz0}
\end{center}
\end{figure}

{\bf Case 1: $-\hat{x} \, , +\hat{y} \, , \Delta \hat{z} = 0$ direction (see Figure~\ref{fig:3drules_xmypz0})}
In this case, the ``2'' signal starts at site $i$ and propagates in the
$+\hat{y}$ direction according to some of the basic rules (8 and 2).
Since $j$ lies in the $-\hat{x}$ direction, the ``3''
signal starts at site $j$ (rule 19) and propagates in the $+\hat{x}$ direction (rule 17).
These two signals meet at their intersection point, and the interaction is completed by a new
type of ``turning'' tensor given by rule 18.

\begin{figure}[ht]
\begin{center}
\begin{tabular}{c}
\includegraphics[width=0.3\textwidth]{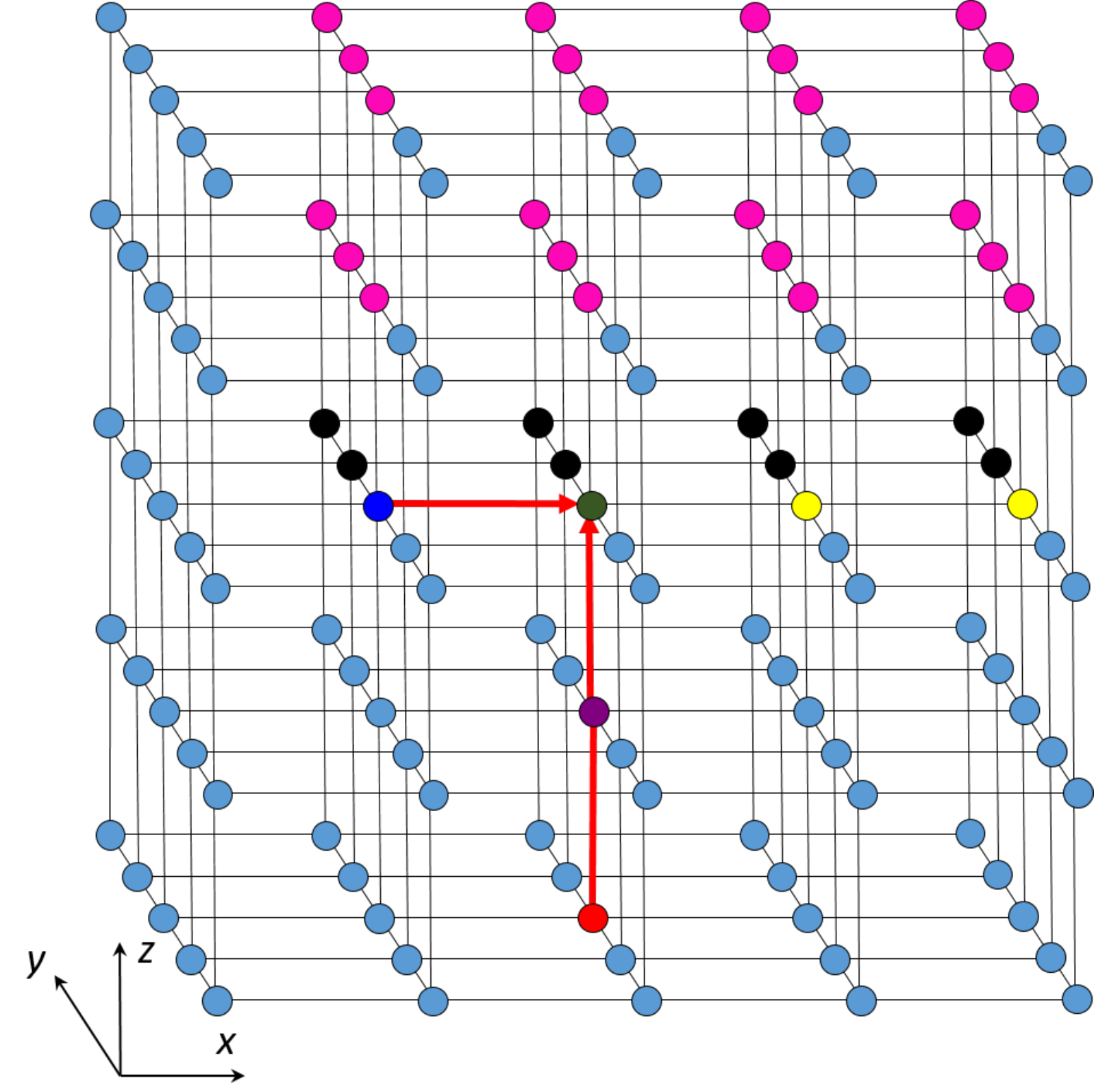}\\
(a) \\
\includegraphics[width=0.3\textwidth]{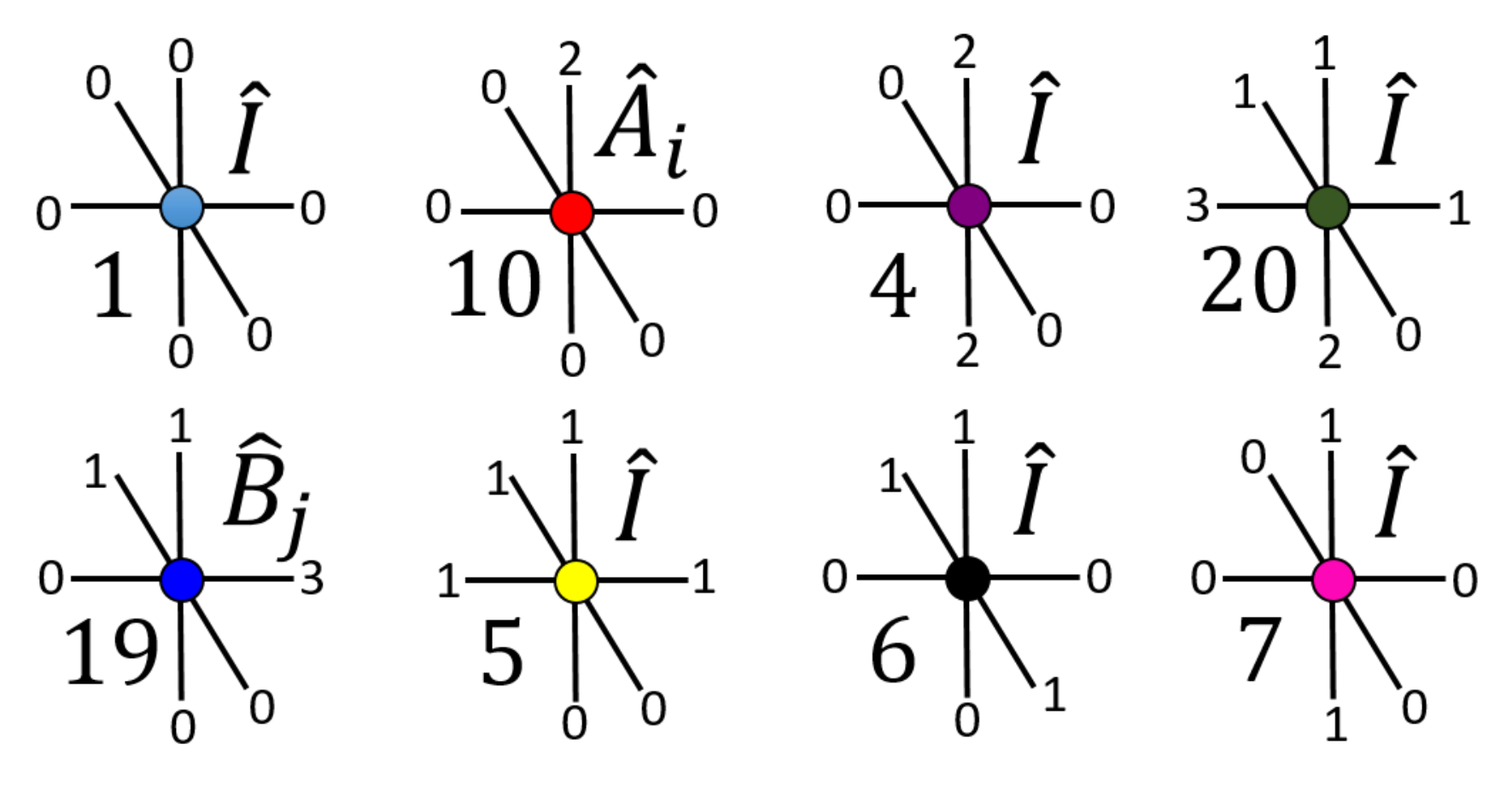}\\
(b) \\
\end{tabular}
\caption{Case 2: $\{ -\hat{x}, \Delta \hat{y} = 0,  +\hat{z} \}$.}
\label{fig:3drules_xmy0zp}
\end{center}
\end{figure}

{\bf Case 2: $-\hat{x} \, , \Delta \hat{y} = 0 \, , +\hat{z}$ direction (see Figure~\ref{fig:3drules_xmy0zp})}
In this case, the ``2'' signal starts at site $i$ and propagates in the $+\hat{z}$ direction
according to basic rules 10 and 4. Since $j$ lies in the  $-\hat{x}$ direction, the ``3''
signal starts at site $j$ (rule 19) and propagates in the $+\hat{x}$ direction (rule 17).
These two signals meet at their intersection point, and the interaction is completed by a new
type of ``turning'' tensor given by rule 20.

\begin{figure}[ht]
\begin{center}
\begin{tabular}{c}
\includegraphics[width=0.3\textwidth]{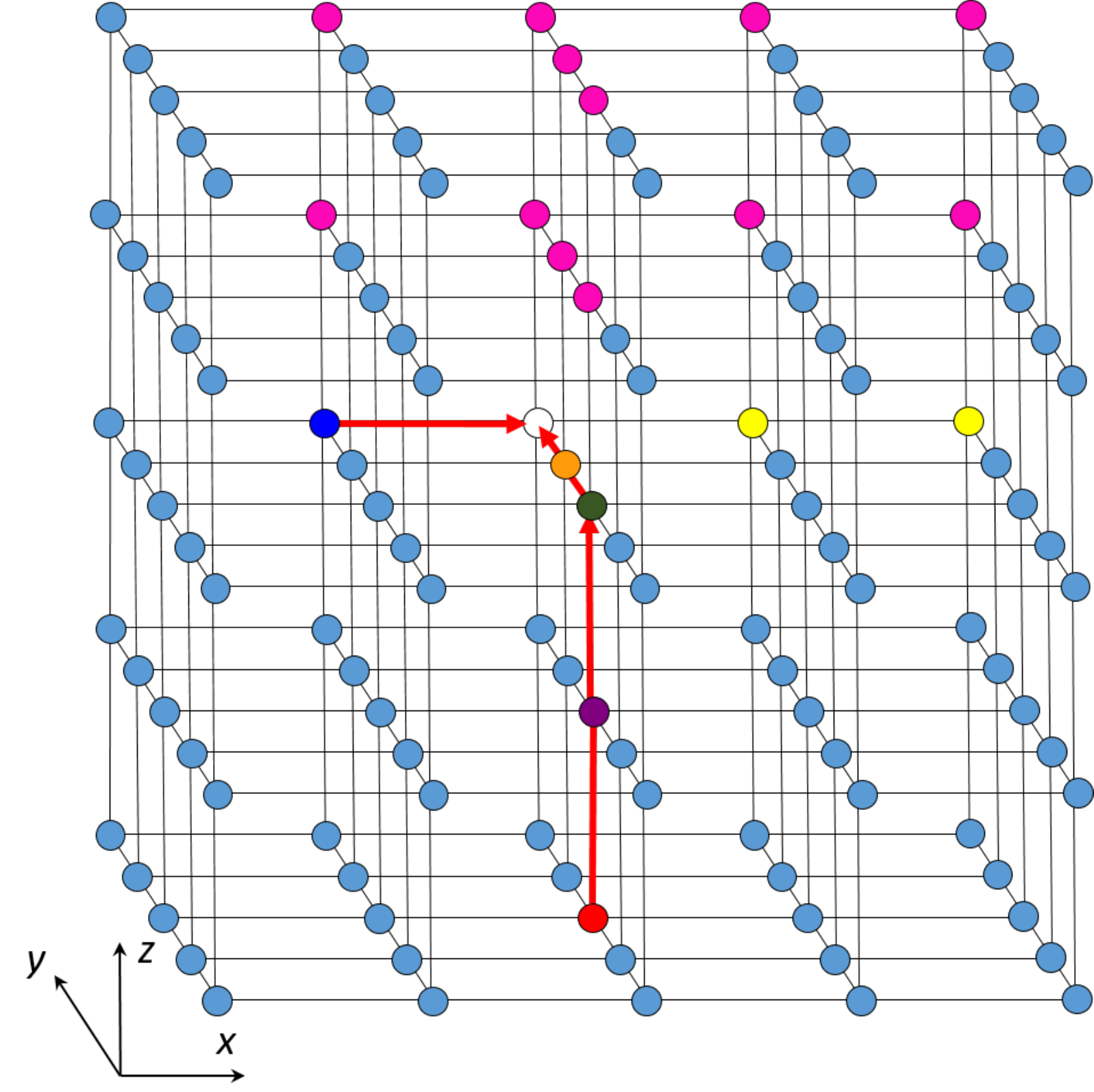}\\
(a) \\
\includegraphics[width=0.3\textwidth]{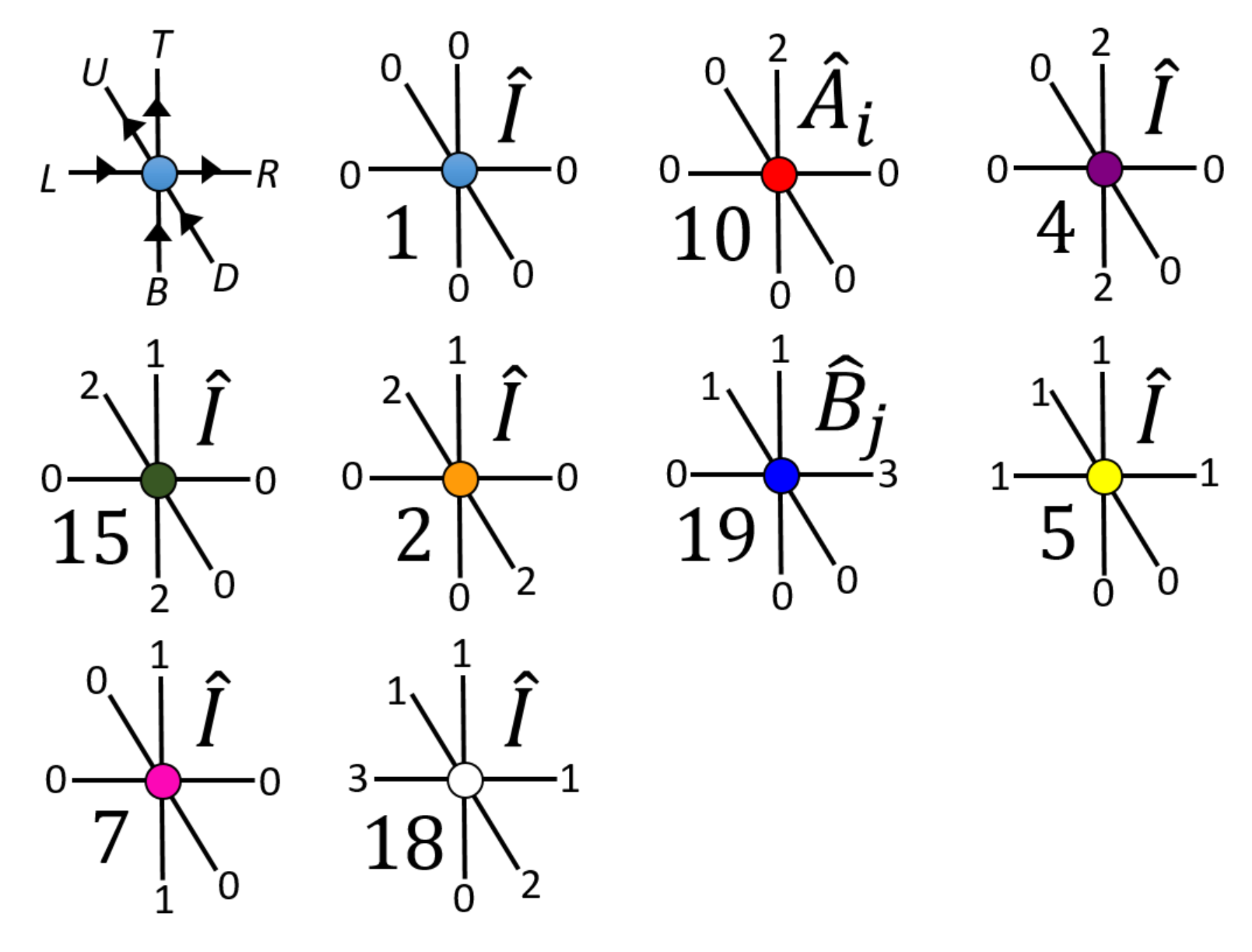}\\
(b) \\
\end{tabular}
\caption{Case 3: $\{ -\hat{x}, +\hat{y}, +\hat{z} \}$.}
\label{fig:3drules_xmypzp}
\end{center}
\end{figure}

{\bf Case 3: $-\hat{x} \, , +\hat{y} \, , +\hat{z}$ direction (see Figure~\ref{fig:3drules_xmypzp})}
In this case, the ``2'' signal starts at site $i$ and first propagates in the
$+\hat{z}$ direction (rules 10 and 4). It then ``turns'' to the $+\hat{y}$ direction (rule 15)
and propagates (rule 2).
Since $j$ lies in the $-\hat{x}$ direction, the ``3''
signal starts at site $j$ (rule 19) and propagates in the $+\hat{x}$ direction (rule 17).
These two signals meet at their intersection point, and the interaction is completed by the ``turning'' tensor given in rule 18.

\begin{figure}[ht]
\begin{center}
\begin{tabular}{c}
\includegraphics[width=0.3\textwidth]{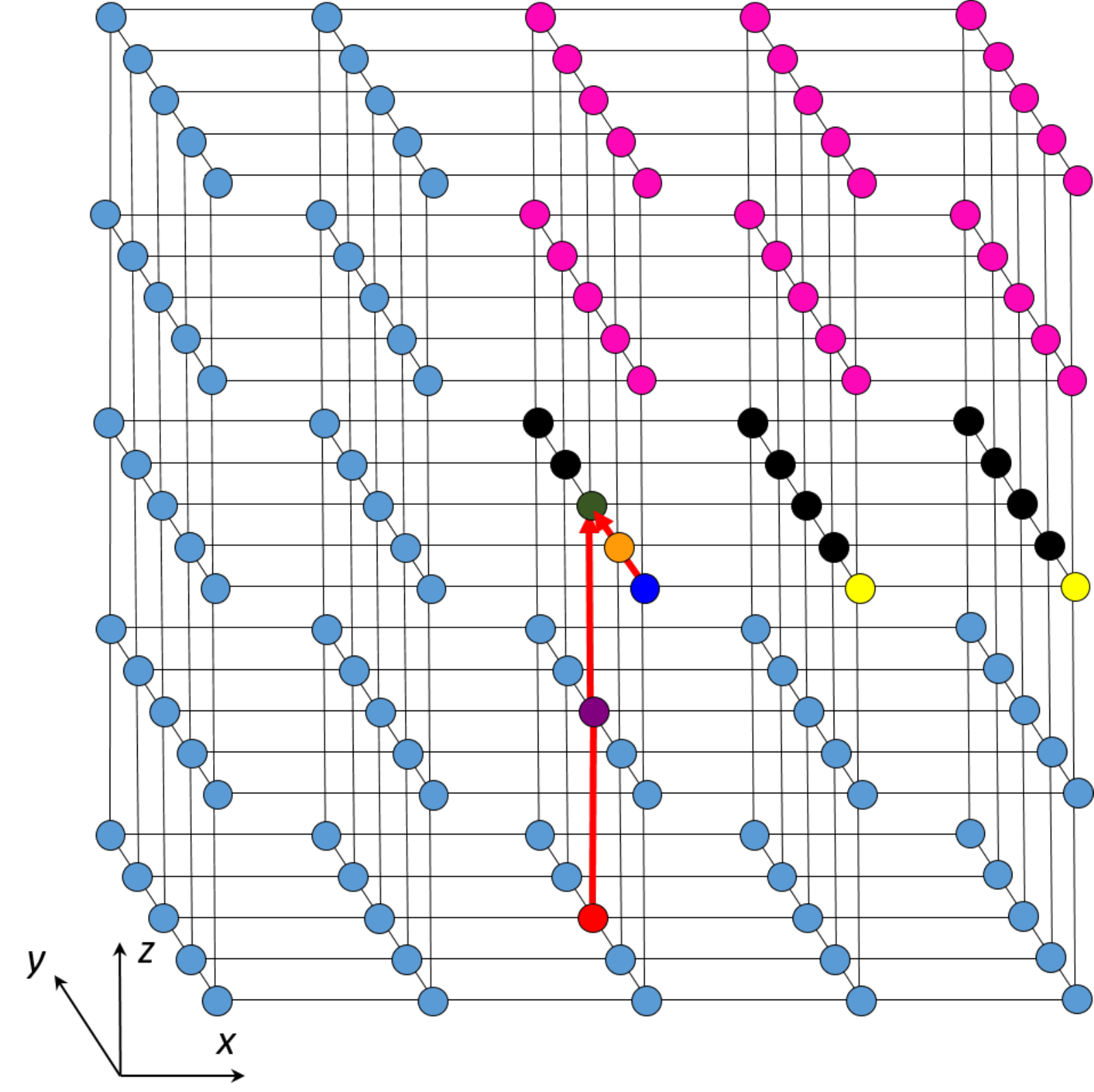}\\
(a) \\
\includegraphics[width=0.3\textwidth]{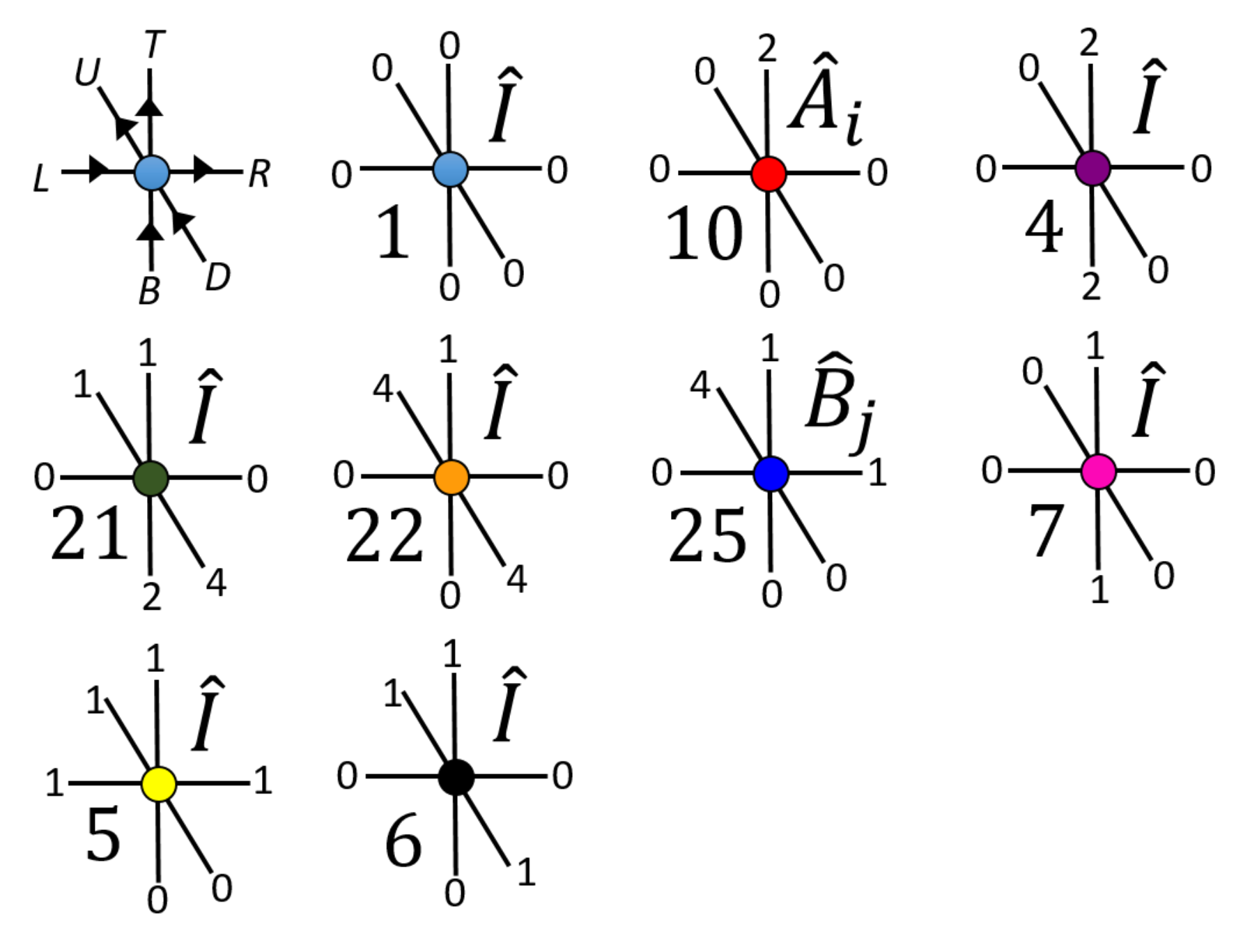}\\
(b) \\
\end{tabular}
\caption{Case 4: $\{ \Delta \hat{x} = 0, -\hat{y}, +\hat{z} \}$.}
\label{fig:3drules_x0ymzp}
\end{center}
\end{figure}

{\bf Case 4: $ \Delta \hat{x} = 0 \, , -\hat{y} \, , +\hat{z}$ direction (see Figure~\ref{fig:3drules_x0ymzp})}
In this case, the ``2'' signal starts at site $i$ and propagates in the $+\hat{z}$ direction
according to basic rules 10 and 4. Since $j$ lies in the  $-\hat{y}$ direction, the ``4''
signal starts at site $j$ (rule 25) and propagates in the $+\hat{y}$ direction (rule 22).
These two signals meet at their intersection point, and the interaction is completed by a new
type of ``turning'' tensor given by rule 21.

\begin{figure}[ht]
\begin{center}
\begin{tabular}{c}
\includegraphics[width=0.3\textwidth]{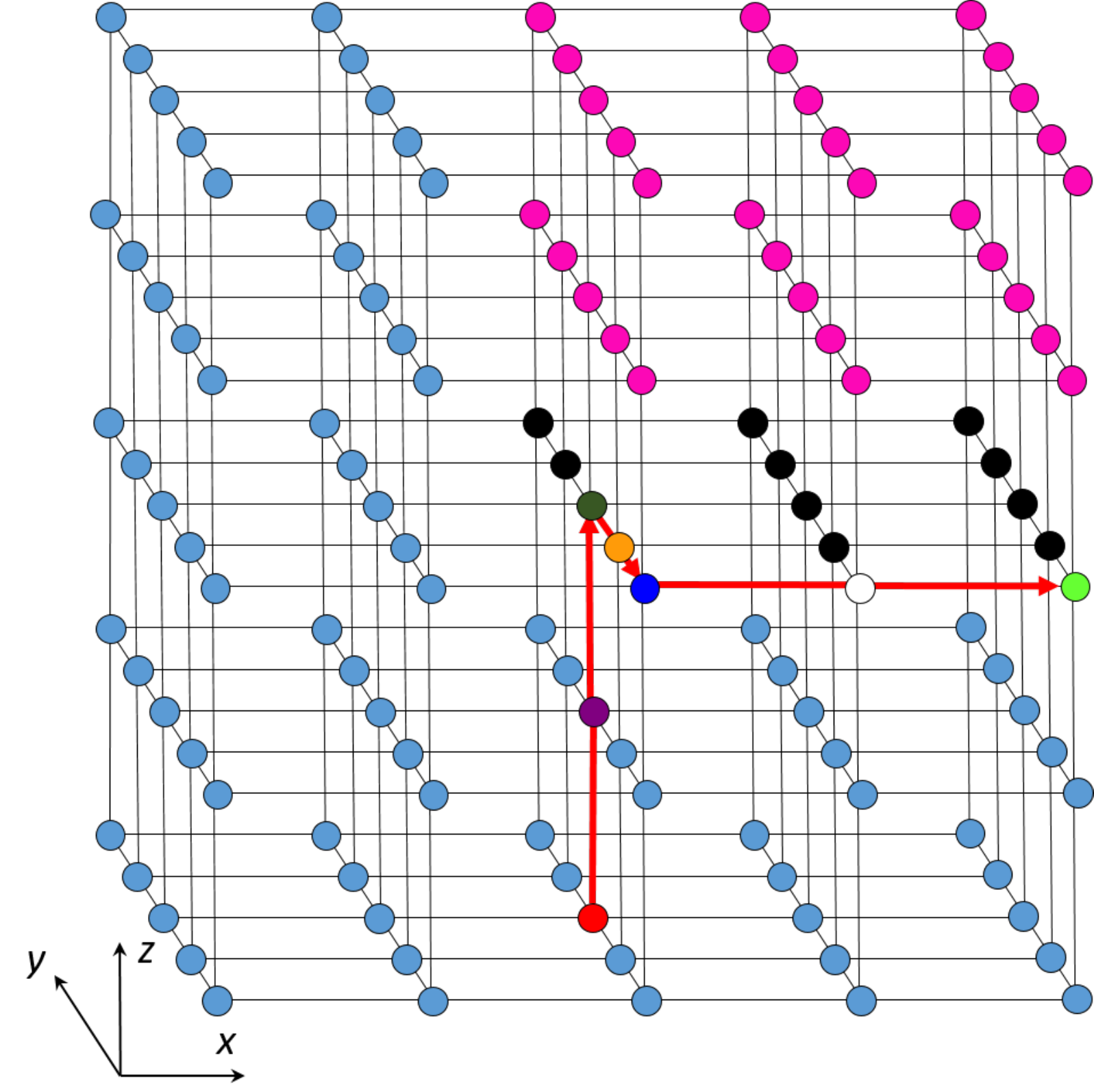}\\
(a) \\
\includegraphics[width=0.3\textwidth]{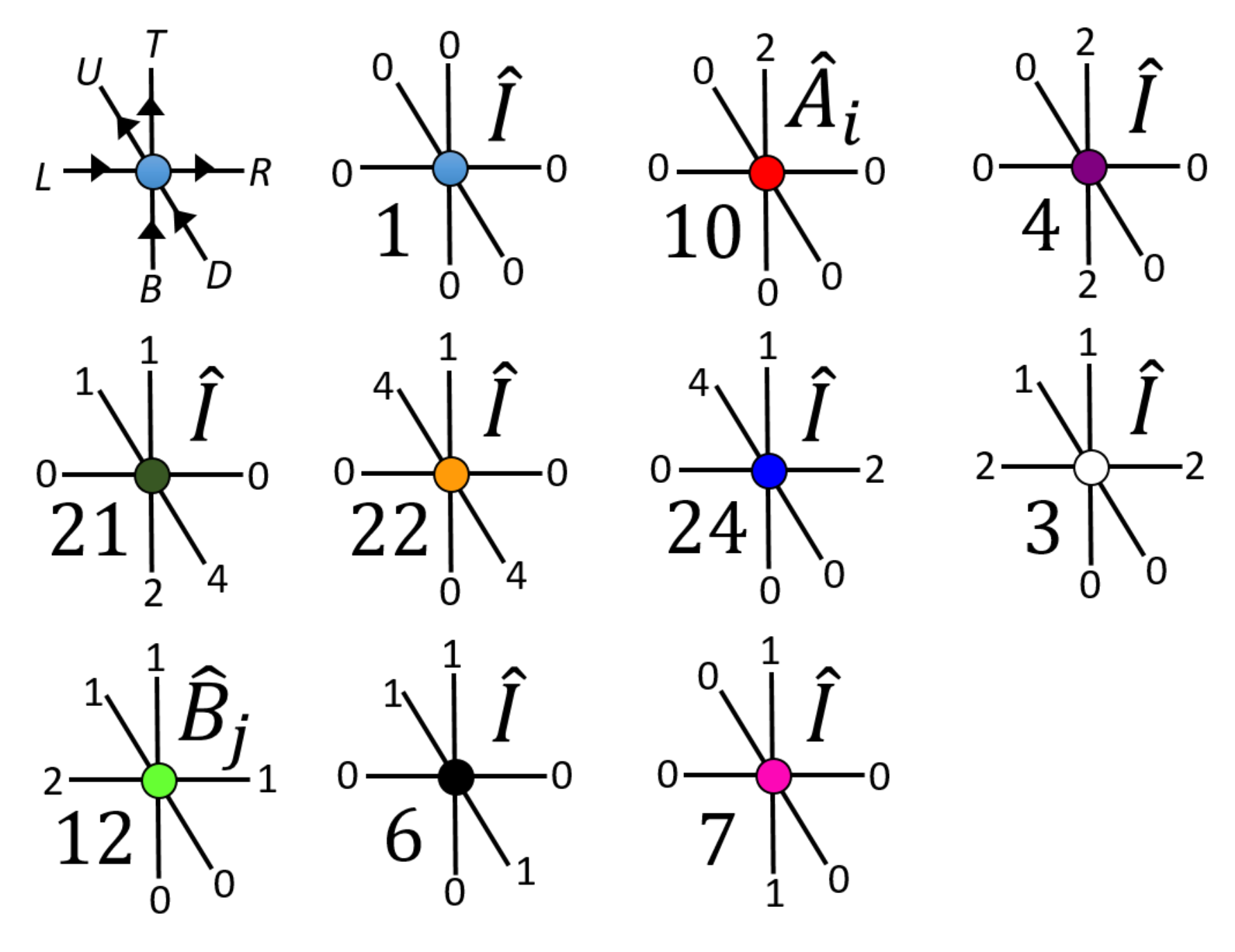}\\
(b) \\
\end{tabular}
\caption{Case 5: $\{ +\hat{x}=0, -\hat{y}, +\hat{z} \}$.}
\label{fig:3drules_xpymzp}
\end{center}
\end{figure}

{\bf Case 5: $ +\hat{x} \, , -\hat{y} \, , +\hat{z}$ direction (see Figure~\ref{fig:3drules_xpymzp})}
Since our convention is to propagate the ``interaction path'' first in the $\hat{z}$
direction, then $\hat{y}$, then $\hat{x}$, this case is a bit less intuituve than the
preceding ones. In our previous analysis, we have pictured the ``4'' as originating from
site $j$ and propagating in the $+\hat{y}$ direction. However, the present case is more easily
understood if we adopt a different (but equivalent) picture in which the ``4'' signal is
a special component of the interaction signal propagating from site $i$ to site $j$ which
is allowed to travel in the $-\hat{y}$ direction, against the directed edge.

Using this new picture, we start as usual with the ``2'' signal originating at site $i$ and
propagating in the $+\hat{z}$ direction according to basic rules 10 and 4. Next, the signal
``turns'' from the $+\hat{z}$ direction to the $-\hat{y}$ direction, becoming a ``4''
(rule 21, as in the previous case). The ``4'' then propagates in the $-\hat{y}$ direction
(rule 22, as above). Finally, the ``interaction signal'' must turn and travel in the $+\hat{x}$
direction and end at site $j$. Since the ``2'' can already go in the $+\hat{x}$ direction and
terminate at $j$ according to basic rules 3 and 12, it can be reused instead of introducing
additional rules. Thus, the ``4'' propagating along $-\hat{y}$ ``turns'' to $+\hat{x}$ and
becomes a ``2'' again according to rule 24, and then basic rules 3 and 12 complete the
interaction.

\begin{figure}[ht]
\begin{center}
\begin{tabular}{c}
\includegraphics[width=0.3\textwidth]{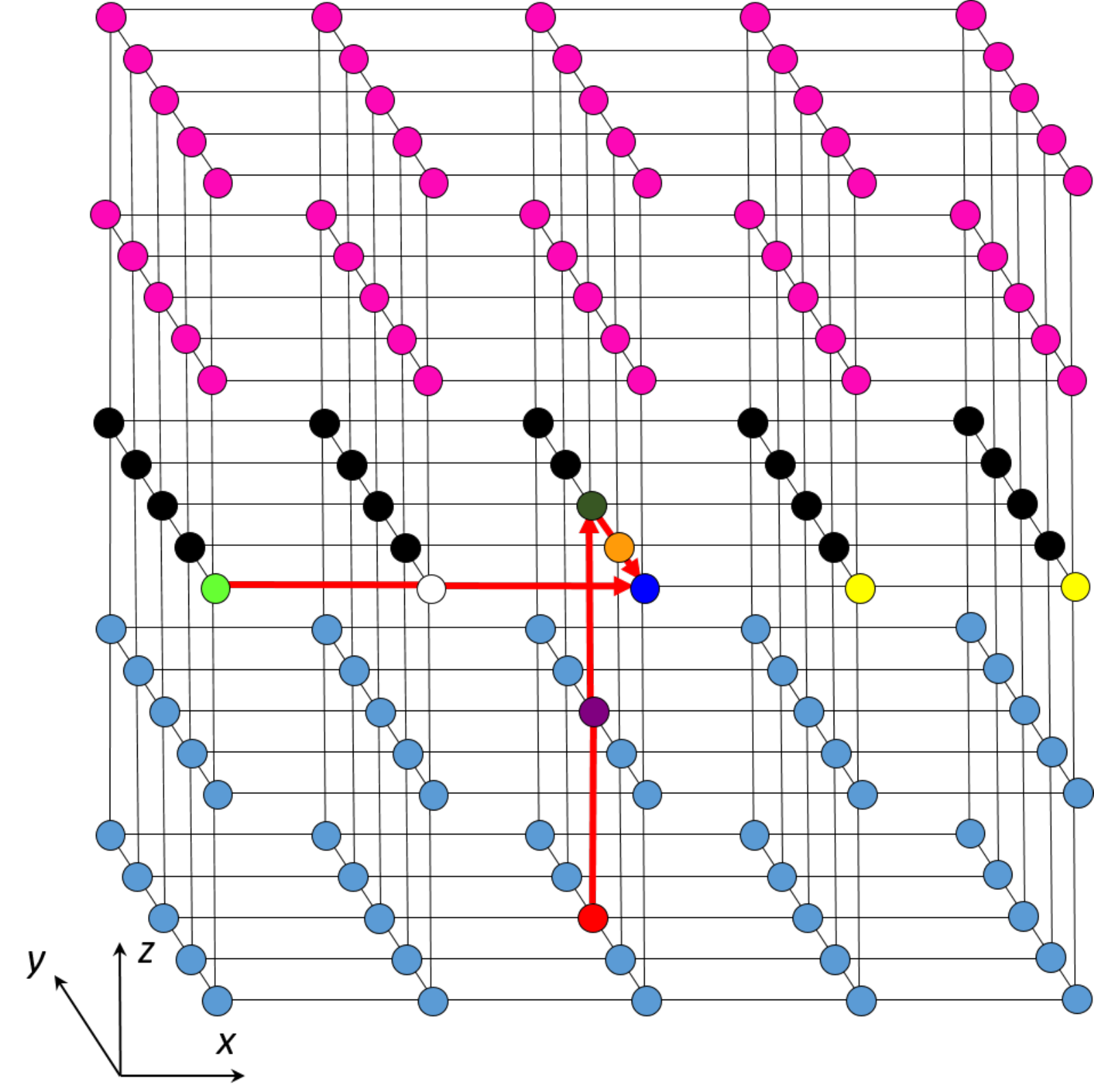}\\
(a) \\
\includegraphics[width=0.3\textwidth]{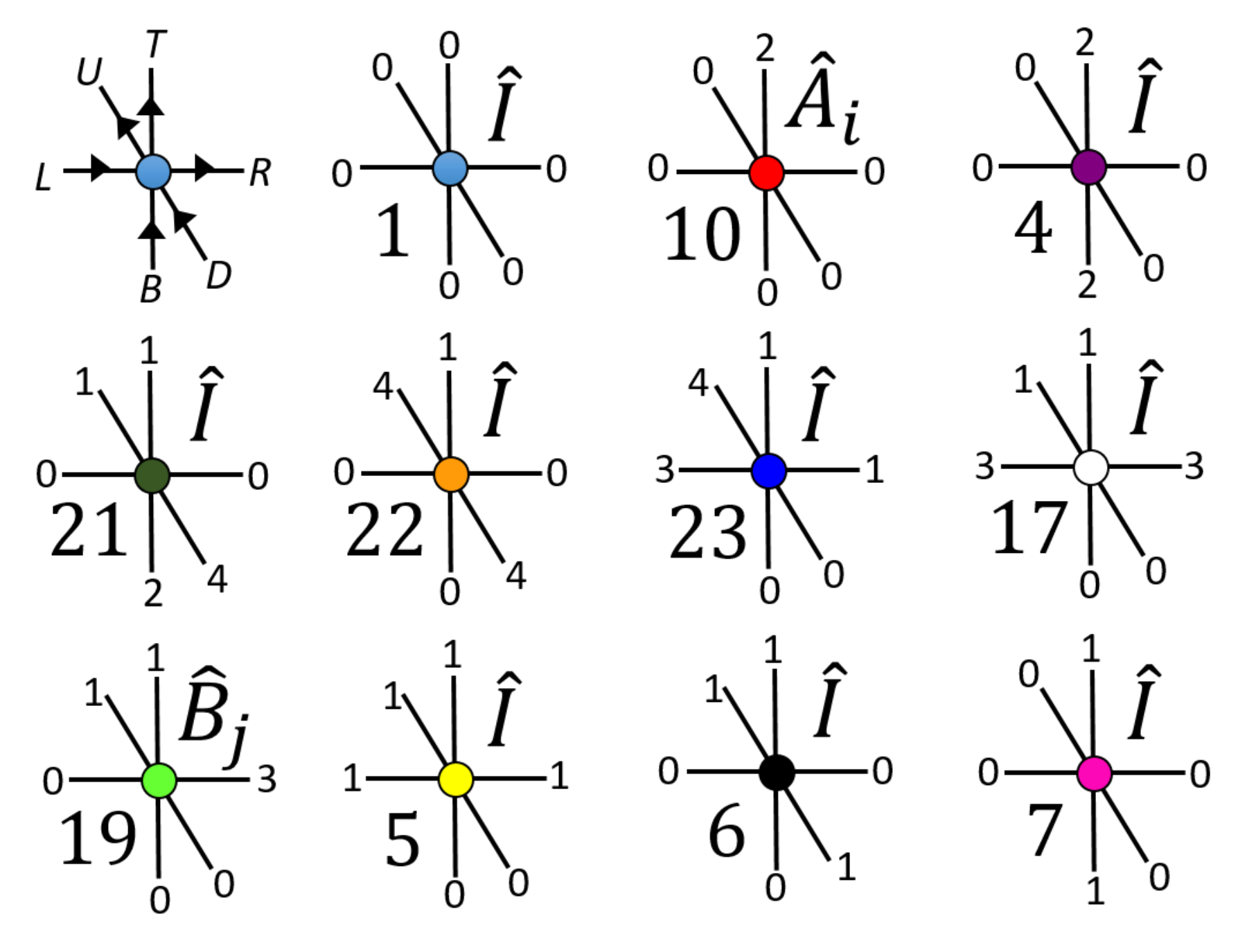}\\
(b) \\
\end{tabular}
\caption{Case 6: $\{ -\hat{x}, -\hat{y}, +\hat{z} \}$.}
\label{fig:3drules_xmymzp}
\end{center}
\end{figure}

{\bf Case 6: $-\hat{x} \, , -\hat{y} \, , +\hat{z}$ direction (see Figure~\ref{fig:3drules_xmymzp})}
In this final case, we combine the two pictures for the ``3'' and ``4'' signals used in
previous cases. First, the ``2'' signal starts at site $i$ and propagates in the $+\hat{z}$ direction
according to basic rules 10 and 4. Next, the signal
``turns'' from the $+\hat{z}$ direction to the $-\hat{y}$ direction, becoming a ``4''
(rule 21) and then propagates in the $-\hat{y}$ direction
(rule 22). Since $j$ lies in the $-\hat{x}$ direction, the ``3''
signal starts at site $j$ (rule 19) and propagates in the $+\hat{x}$ direction (rule 17).
The ``3'' and ``4'' then meet at their intersection point, and the interaction is completed
by a new type of ``turning'' tensor given by rule 23.

{\bf Final rule:} Up to this point, all the rules in Table~\ref{tab:3drules} have been utilized except rule 26.
This is a special rule that only applies to the tensor in the top right corner of the top plane
of the network, where the finite state machine terminates. This rule is included to disallow
the state of the machine where all tensors have virtual index
values $(0,0,0,0,0,0)$ and a spurious 1 is added to Eq.~\eqref{eqn:AiBj}
so that the final operator is $1+\sum_{i < j} \hat{A}_i \hat{B}_j$ instead of the target $\sum_{i < j} \hat{A}_i \hat{B}_j$.

\clearpage
\bibliographystyle{apsrev4-1}
\bibliography{references}

\end{document}